\documentclass{IEEEtaes}

\usepackage[T1]{fontenc}
\usepackage{graphicx}
\usepackage{cite}
\usepackage{caption}
\usepackage{subfigure}
\usepackage{epstopdf}

\usepackage{overpic}
\usepackage{amssymb}
\usepackage{amsmath}
\usepackage{amsthm}
\usepackage{array}
\usepackage{color}
\usepackage{url}

\usepackage{acronym}
\usepackage{algorithmic}
\usepackage{algorithm}

\usepackage{multirow}

\IEEEoverridecommandlockouts

\theoremstyle{plain}

\newcommand{\figuresname}{Figs.}
\newcommand{\sectionname}{Sect.}

\renewcommand{\tablename}{Table}

\acrodef{ADC}{analog-to-digital conversion}
\acrodef{B5G}{beyond-5G}
\acrodef{CS}{central satellite}
\acrodef{DAC}{digital-to-analog conversion}
\acrodef{DRA}{direct radiating antenna}
\acrodef{DC}{direct current}
\acrodef{DO}{digital-to-optical}
\acrodef{OD}{optical-to-digital}
\acrodef{eMBB}{enhanced mobile broadband}
\acrodef{FoA}{formation of arrays}
\acrodef{GEO}{geostationary earth orbit}
\acrodef{KPI}{key performance indicator}
\acrodef{LEO}{low earth orbit}
\acrodef{MAI}{multiple access interference}
\acrodef{MEO}{medium earth orbit}
\acrodef{mMTC}{massive machine-type communications}
\acrodef{mMIMO}{massive multiple-input multiple-output}
\acrodef{eMTC}{enhanced machine-type communications}
\acrodef{NTN}{non-terrestrial network}
\acrodef{MMSE}{minimum mean square error}
\acrodef{VHTS}{very high throughput satellite}
\acrodef{URLLC}{ultra-reliable low-latency communications}
\acrodef{ECEF}{earth-centered earth-fixed}
\acrodef{GPS}{global positioning system}
\acrodef{HTFS}{high throughput fractionated satellite}
\acrodef{UHF}{ultra-high frequency}
\acrodef{FF}{formation flying}
\acrodef{MIMO}{multiple input multiple output}
\acrodef{GNC}{guidance navigation and control}
\acrodef{GNSS}{global navigation satellite system}
\acrodef{CSS}{central sub-satellite}
\acrodef{SS}{sub-satellite}
\acrodef{GW}{gateway}
\acrodef{ISL}{inter-satellite link}
\acrodef{BFN}{beamforming/feeding network}
\acrodef{SNR}{signal-to-noise ratio}
\acrodef{SIR}{signal-to-interference ratio}
\acrodef{SINR}{signal-to-interference-plus-noise ratio}
\acrodef{RF}{radio frequency}
\acrodef{R-GEO}{regional \acs{GEO}}
\acrodef{EIRP}{effective isotropic radiated power}
\acrodef{UT}{user terminal}
\acrodef{UE}{user equipment}
\acrodef{TDM}{time division multiplexing}
\acrodef{FDM}{frequency division multiplexing}
\acrodef{RRM}{radio resource management}
\acrodef{SG}{solar generator}
\acrodef{SSPA}{solid-state power amplifier}
\acrodef{AWGN}{additive white Gaussian noise}
\acrodef{LoS}{line-of-sight}
\acrodef{PHY}{physical-layer}
\acrodef{BLER}{block error rate}
\acrodef{DL}{downlink}
\acrodef{UL}{downlink}
\acrodef{PSD}{power spectral density}
\acrodef{SE}{spectral efficiency}
\acrodef{WDM}{wavelength division multiplexing}
\acrodef{ROI}{region of interest}

\newcommand{\verIndex}{\ensuremath{n}}
\newcommand{\horIndex}{\ensuremath{m}}

\newcommand{\verElements}{\expandafter\MakeUppercase\expandafter{\verIndex}}
\newcommand{\horElements}{\expandafter\MakeUppercase\expandafter{\horIndex}}

\def\CN{\mathcal{N}_{\mathbb{C}}} 
\def\imagunit{\mathsf{j}} 

\begin{document}
\title{Formation-of-Arrays Antenna Technology for High-Throughput Mobile Non-Terrestrial Networks}
\author{
Giacomo Bacci$^{1}$, Riccardo De Gaudenzi$^2$,  Marco Luise$^{1,3}$, Luca Sanguinetti$^{1}$, Elena Sebastiani$^{1,4}$\\
\IEEEauthorblockA{\small{$^1$Dipartimento Ingegneria dell'Informazione, University of Pisa, Italy (\{giacomo.bacci, marco.luise, luca.sanguinetti\}@unipi.it)} \\
  \small{$^2$European Space Agency, Keplerlaan 1, 2200 AG Noordwijk, The Netherlands (rdegaude@gmail.com)} \\
  \small{$^3$CNIT / University of Florence, Italy} \\
  \small{$^4$WISER srl, Livorno, Italy (elena.sebastiani@wiser.it)}}
}
%

\receiveddate{Manuscript received August 12, 2022; revised October 10, 2022, and December 14, 2022; accepted February 7, 2023. This work was partially supported by the European Union under the Italian National Recovery and Resilience Plan (NRRP) of NextGenerationEU, partnership on ``Telecommunications of the Future'' (PE00000001 - program ``RESTART'').}

\markboth{G. BACCI ET AL.}{FOA ANTENNA TECHNOLOGY FOR HIGH-THROUGHPUT MOBILE NON-TERRESTRIAL NETWORKS}
\maketitle
\begin{abstract}
Effective integration of terrestrial and non-terrestrial segments is one of the key research avenues in the design of current and future wireless communication networks. To this aim, modern communication-satellite constellations intend to attain sufficiently high throughput in terms of bit rate per unit area on the ground by rather aggressive patterns of spatial frequency reuse. This goal calls for on-board narrow-beam antennas, whose size turns out to be in many cases incompatible with the size/mass and accommodation constraints of the hosting satellite. This paper investigates the attainable performance of \emph{large distributed arrays of antennas} implemented as the ensemble of a few to many simpler sub-antennas of smaller sizes, carried by one (small) satellite each. The sub-antennas can in their turn be implemented like (regular) 2D arrays of simple radiating elements, realizing an overall (distributed) antenna architecture that we call ``\acl{FoA}'' (\acs{FoA}). The satellites that implement this radiating architecture need to be relatively close to each other and constitute a \emph{formation} of flying objects, to be coordinated and controlled as a whole. In this paper, we develop a theoretical analysis of an \acs{FoA} antenna, and we show how to take advantage of this new technology to improve network throughput in a multi-beam S-band mobile communication network  with low-earth or geostationary orbiting satellites directly providing 5G-like communication services to hand-held user terminals. 
\end{abstract}
\IEEEpeerreviewmaketitle
%
%
\section{Introduction and Motivation}
\label{Sec:Introduction_and_Motivation}
The community of 5G and \ac{B5G} networks development and standardization expresses a growing interest in integrating a \ac{NTN} segment in current (5G) and next generation (\ac{B5G}) terrestrial networks for seamless mobile communications \cite{Rinaldi2020}. Historically, satellites proved to be an attractive complement to the terrestrial infrastructure for covering scarcely populated regions (i.e., where terrestrial infrastructure is not economically viable), or where no ground infrastructure can be deployed altogether (e.g., over the sea). Speaking of 5G, \ac{GEO} satellites with their wide and stationary coverage area are ideal for supporting time-noncritical applications, while \ac{LEO}-satellite constellations are more suited to meet the fundamental 5G \ac{URLLC} / \ac{eMBB} requirements. \Ac{mMTC} can be supported by both \ac{GEO} and \ac{LEO} constellations. For this reason, work is in progress in the framework of 5G 3GPP standardization to include \acp{NTN} starting from release $17$ of the standard \cite{3GPP_NTN}.

True integration of \acp{NTN} with cellular networks also calls for the possibility to adopt a universal user terminal not  different  from what is currently used in conventional cellular networks in terms of weight, power consumption and form factor/antenna. Whatever the kind of constellation and/or use case one wishes to address, successful integration needs from this standpoint a major boost (i.e., orders of magnitude improvement) in terms of throughput and/or served user density with respect to the current state of the art of satellite technology \cite{DeGaudenzi5G2022}.
The required bitrate per unit area increase calls for a ultra-high gain multi-beam satellite antenna with a rather ``aggressive'' pattern of spatial frequency reuse, akin to what is done by terrestrial cellular networks.

In this respect, our starting point is the initial consideration of a broadband 4G-like satellite mobile service featuring conventional hand-held terminals with a best-case outdoor line-of-sight link and modest inter-beam interference. The resulting space segment payload requirements are shown in \tablename~\ref{tab:Example_GEO_LEO_sizing} for a \ac{LEO} and for a \ac{R-GEO} satellite, both equipped with an active planar-array antenna.\footnote{The assumptions that appear in \tablename~\ref{tab:Example_GEO_LEO_sizing} will be thoroughly justified in the next sections, and introduced here with no further comment as a motivation of our work.}
From this preliminary, simplified use cases  we see  that a modest aggregate throughput of $10.7$ Mbps in a single beam for the \ac{R-GEO} satellite (to be shared by all of the users within the relative $14$-km beam radius footprint) requires a very challenging (not to say impossible to deploy altogether) active array antenna of $14762$ m$^2$ and $53$ dBW of on-board \ac{RF} power. The same $10.7$-Mbps beam throughput over a beam size of $3.51$ km requires a (much) more reasonable \ac{LEO} active array of $71$ m$^2$ and an \ac{RF} power of $28$ dBW.

Above, we have implicitly assumed that the preferred technology to implement a high-gain flexible-coverage multi-beam antenna is represented by (phased) arrays, a technology that in a small scale can be considered consolidated by now. However, the issue of how to accommodate, launch, and deploy a very large phased array like the ones envisaged above is still a challenge, especially if a further low-cost constraint is added. Deployable antenna reflectors in excess of $18$ meters have been adopted by \ac{GEO} satellites for mobile applications \cite{Semler2010}, but their size is anyway insufficient (\tablename~\ref{tab:Example_GEO_LEO_sizing}) to provide 4G-like (let alone 5G) services to hand-held devices. Current \ac{LEO} satellites in megaconstellations feature antennas size not exceeding $1$-$2$ meters, which is barely sufficient for 2G/3G type of mobile services only. This explains the recent interest to consider \emph{distributed} antennas across a swarm of satellites in a \ac{FF} configuration \cite{Bekey2006,AST_patent2020,Zhang2011,Deng2021,Delamotte2021}.

\begin{table}[t]
\centering
\scriptsize{
  \begin{tabular}{l c c c}\hline
    \multirow{2}{*}{\emph{Parameter}}  & \emph{\acs{R-GEO} case} & \emph{\acs{LEO} case} & \multirow{2}{*}{\emph{Unit}} \\
                                         & \emph{value}            & \emph{value}          & \\\hline
    Carrier frequency                    &      $2.2$                &    $2.2$                & GHz   \\
    System allocated bandwidth           &      $60$                 &    $60$                 & MHz   \\
    Orbital height                       &      $35,870$             &    $550$                & km    \\
    Total satellite \acs{RF} power       &      $53.3$               &    $28.0$               & dBW    \\
    Satellite antenna array area         &      $14,762$             &    $70.8$               & m$^2$\\
    Satellite EIRP/beam                  &      $83.0$               &    $43.2$               & dBW   \\
    User satellite elevation angle       &      $60$                 &    $40$                 & degrees\\
    \Ac{SIR}                             &      $6.6$                &    $3.4$                & dB    \\
    User terminal antenna $G/T$          &      $-31.6$              &    $-31.6$              & dB/K  \\
    Mobile channel average shadowing     &      $4.0$                &    $4.0$                & dB    \\
    Body obstruction losses              &      $3.0$                &    $3.0$                & dB    \\
    Rician fading  extra margin          &      $3.0$                &    $3.0$                & dB    \\
    Approximate beam radius (ground)     &      $14.2$               &    $3.1$                & km    \\
    Single-beam throughput               &      $10.7$               &    $10.7$               & Mbps  \\
    Number of active beams               &      $6777$               &    $54$                 &       \\
    Aggregate throughput                 &      $72.7$               &    $1.8$                & Gbps  \\\hline
  \end{tabular}
}
\caption{Example of downlink system sizing for \acs{R-GEO} and \acs{LEO} cases.}
\label{tab:Example_GEO_LEO_sizing}
\end{table}

The concept of using a \ac{FF} approach to form extremely narrow-beam distributed antennas has first appeared in the field of radio-astronomy and Earth observation, and was later on introduced in \cite{Bekey2006} for telecommunication satellites, where the author proposed to combine a number of \ac{LEO} pico-satellites with a single (simple) radiating element to obtain a distributed array with a narrow on-ground beam. A central ``master'' satellite feeding the pico-satellite swarm takes care of the required beam forming, and a \ac{GPS} receiver allows each pico-satellite to determine its own position, and thus to compute its own required phase delay. Unfortunately, no numerical results concerning antenna radiation are provided in the paper.

Similarly, \cite{AST_patent2018} introduces the concept of a \ac{HTFS} system. The key idea is to distribute the functional capabilities of a conventional single telecommunication spacecraft across a multitude of many small satellites. Such satellites are combined in an accurate \ac{FF} mode, and connected to a central command and relay satellite, so that the distributed architecture allows to regulate in a modular fashion the global satellite antenna aperture. As an alternative, the same company has patented the concept of a very large deployable active modular phased array \cite{AST_patent2020}, currently being built with a target size of $30\times30\,\textrm{m}^2$ for the AST Space Mobile constellation operating in the \ac{UHF} band. In this technology, the sub-arrays are interconnected through mechanical hinges keeping them in a fixed spatial multi-array configuration. Although in this paper we tackle the more general problem of a composite array created by the composition of \ac{FF} satellite sub-arrays, the investigation on their geometrical optimization is also applicable to the case of deployable phased arrays like the one described in \cite{AST_patent2020,Couchman_deployable_phased_array}. For antennas larger than this (as may be requested by a GEO satellite), one can also envisage in-space building of a composite array that is launched and first deployed like a \ac{FoA}, then mechanically stabilized by hinging each array to the neighboring ones. 

The idea of distributed yet coordinated communication satellites is pursued in \cite{Zhang2011}, that provides an extension of the 3GPP \ac{NTN} mobile channel \ac{MIMO} model to include the presence of a constellation of \ac{LEO} satellites in \ac{FF} combined with an on ground array. More recently, the authors in \cite{Deng2021} derived the theoretical \ac{MIMO} capacity for an ultra-dense \ac{LEO} network with cooperating multiple Earth terminals and \ac{FF} satellites. The satellite formation on one side, as well as the Earth terminals array on the other, are seen as two single entities, so that the total point-to-point capacity is evaluated.

In this paper, we concentrate on the antenna system that is structured as a virtual large array of smaller antennas, each one on-board a small satellite. The sub-antennas are implemented like (regular) 2D phased arrays composed by a reduced number of radiating elements, realizing an overall (larger, distributed) antenna architecture: the FoA. The satellites that are grouped to realize this distributed radiating architecture need to be relatively close to each other and constitute a formation of flying objects, to be coordinated and controlled as a whole. The advantage of our \ac{FoA} approach is pretty clear: realizing the desired, very narrow beam radiation pattern without recurring to a costly and problematic very large deployable active antenna technology. The disadvantage is clear as well: the formation of satellites has to be stable enough in both the orbital and the radio frequency sense.

A distributed antenna like the \ac{FoA} (that can be considered a \emph{sparse} array, i.e., a very large array with inner gaps) can be as large as needed (e.g., even hundred meters in diameter in case of a \ac{GEO}), making a very large base available to create ultra-narrow beams. Of course, due to the array sparsity, the radiation pattern will not be the same as that of an equivalent-size non-sparse (conventional) 2D array. This is a challenge we wish to address in this paper: finding a sufficiently ``good'' geometric configuration of the satellite formation and of the single (sub)antennas carried by each spacecraft originating a sufficiently ``good'' radiation pattern of the \ac{FoA}. Such aspects will be addressed and discussed in \sectionname~\ref{sec:model}, before some sample \ac{FoA} configurations are presented, namely, a \ac{GEO} regional system and a global \ac{LEO} constellation, that will be described in detail in \sectionname~\ref{sec:system}. Some performance results in terms of communications throughput will be derived in \sectionname~\ref{sec:results}, and some practical system implementation issues are discussed in \sectionname~\ref{sec:implementation}. The customary concluding remarks will be finally presented in \sectionname~\ref{sec:ConcSect}.
%

\section{Model and analysis of the \acl{FoA}} \label{sec:model}

\begin{figure}[t]
  \begin{center}
    \begin{overpic}[width=\columnwidth]{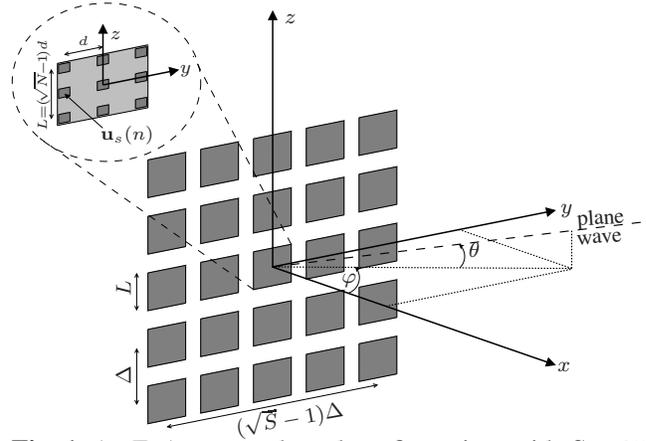}
      \put(86,09){\footnotesize{$x$}}
      \put(86.5,34){\footnotesize{$y$}}
      \put(43,64){\footnotesize{$z$}}
      \put(26.5,56){\scriptsize{$y$}}
      \put(15.5,61.5){\scriptsize{$z$}}
      \put(52,23){\footnotesize{{$\varphi$}}}
      \put(72,27){\footnotesize{{$\theta$}}}
      \put(16.5,21){\rotatebox{90}{\footnotesize{$L$}}}
      \put(16.5,07){\rotatebox{90}{\footnotesize{$\Delta$}}}
      \put(3.5,47){\rotatebox{90}{{\tiny{$L\!\!=\!\!(\!\sqrt{N}\!\!-\!\!1\!)d$}}}}
      \put(10.5,60.5){\tiny{$d$}}
      \put(14,45.5){\scriptsize{$\mathbf{u}_s\!\left(n\right)$}}
      \put(35,-1){\rotatebox{10}{{\footnotesize{$(\sqrt{S}-1)\Delta$}}}}
      \put(89,32){\rotatebox{4}{{\footnotesize{{plane}}}}}
      \put(89,29){\rotatebox{4}{{\footnotesize{{wave}}}}}
    \end{overpic} 
    \caption{An \acs{FoA} square-shaped configuration, with $S=25$ satellites (arrays), located in the $yz$-plane, each hosting $N=9$ radiating elements
      with spacing $d$, also showing a transmitted plane wave with elevation $\theta$ and azimuth $\varphi$.}
    \label{fig:FoA}
  \end{center}
\end{figure}

We assume that the \ac{FoA} is composed by a planar arrangement of $S$ satellites, placed on the $yz$-plane, each equipped with an array of $N$ radiating elements, as shown in \figurename~\ref{fig:FoA}. We use $\mathbf{u}_{s}(n)\in \mathbb{R}^3$ to
indicate the position (with respect to the origin of the reference system) of the $n$-th radiating element within the array hosted by of the $s$-th satellite, $n=1,\ldots,N$, $s=1,\ldots,S$.\footnote{{For simplicity, in the reminder of the paper we will use the expression ``satellite'' when referring to each satellite hosting a sub-array which is part of the \ac{FoA}. When discussing system implementation issues in \sectionname~\ref{sec:implementation}, we will introduce the expression ``array satellite'' to avoid possible ambiguities.}} We also denote with $\varphi$ and $\theta$ the azimuth and elevation angles of a generic location in space, respectively, and we call $\lambda$ the signal carrier wavelength. 
The \emph{wave vector} $\mathbf{k}\left(\varphi,\theta\right)$, defined as
\begin{align}\label{eq:waveVector}
\mathbf{k}(\varphi,\theta)= \frac{2\pi}{\lambda}\left[\cos\theta\cos\varphi, \;\cos\theta\sin\varphi, \;\sin\theta \right]^T,
\end{align}
regulates the phase variation of a plane wave with respect to the three Cartesian coordinates. In particular, {the phase shift of the far-field plane wave generated by a radiating element located at the arbitrary location $\mathbf{u}$ is equal to $\mathbf{k}^T(\varphi,\theta)\mathbf{u}$}. The \emph{array response vector} $\mathbf{a}_{s}(\varphi,\theta)\in{\mathbb{C}^{N}}$ of the $s$th array with $N$ antennas placed at location $\{\mathbf{u}_{s}(n); n=1,\ldots, N\}$ is
\begin{align}\label{eq:arrayVector}
\mathbf{a}_{s}(\varphi,\theta)\in{\mathbb{C}^{N}}= g(\theta)
    \left[e^{\imagunit {\mathbf{k}}^T(\varphi,\theta) \mathbf{u}_{s}(1)},  \ldots \,
          e^{\imagunit {\mathbf{k}}^T(\varphi,\theta) \mathbf{u}_{s}(N)} \right]^{T},
\end{align}
where $g(\theta)$ is the array element radiation pattern (same pattern for each element).
The \emph{global} response vector of the \ac{FoA} is therefore
\begin{align}\label{eq:FoAVector}
\mathbf{a}(\varphi,\theta) \in {\mathbb{C}^{NS}} =
    \left[\mathbf{a}_{1}^T(\varphi,\theta), \ldots,
       \mathbf{a}_{S}^T(\varphi,\theta)
    \right]^{T}.
\end{align}
This model is valid for any arbitrary geometry of the arrays and the \ac{FoA}, which may vary depending on the use case, and the area to be covered. In the remainder of the paper, we will focus on a specific \ac{FoA} configuration, triggered by practical considerations related to formation flying. We assume that each satellite is equipped with a 2D square planar array consisting of $\sqrt{N}$ horizontal rows with $\sqrt{N}$ radiating elements each, such as the one reported in the magnification of \figurename~\ref{fig:FoA}. This choice is mainly due to considerations on the shape of the launcher fairing, which is typically circular, so that a good trade-off between engineering feasibility of the satellite array and electromagnetic properties is represented by a square shape that  fits well the fairing. The radiating elements are uniformly spaced with horizontal and vertical spacing $d$, so that each array size in both the horizontal and vertical direction is equal to $L=(\sqrt{N}-1)d$. Similarly, we assume a square \ac{FoA} of $S$ arrays arranged on a  $\sqrt{S}\times\sqrt{S}$ uniformly spaced grid, with horizontal/vertical spacing $\Delta\ge L$, as depicted in \figurename~\ref{fig:FoA}. We have also tested different array/\ac{FoA} configurations than square (hexagonal, quincunx, etc.) but our results bore little difference from those obtained with the simple square arrangement that we present here.

Under these assumptions, the position $\mathbf{u}_s(n)$ of the $n$-th antenna in the $s$-th array can be written as $\mathbf{u}_s(n) = \mathbf{r}_s + \mathbf{r}_n$, where $\mathbf{r}_s=\left[x_s, y_s, z_s\right]^T$ is the position of the $s$th array's center (the nominal antenna location in the formation), with
\begin{align}
  \label{eq:arrayPosition_x}
  x_s&=0,\\
  \label{eq:arrayPosition_y}
  y_s&=\Delta \left(-\frac{\left(\sqrt{S}-1\right)}{2} +  \textrm{mod} \left(s-1, \sqrt{S}\right)\right),\\
  \label{eq:arrayPosition_z}
  z_s&=\Delta\left(-\frac{\left(\sqrt{S}-1\right)}{2} + \left\lfloor\frac{s-1}{\sqrt{S}}\right\rfloor\right),
\end{align}
whilst $\mathbf{r}_n=\left[x_n, y_n, z_n\right]^T$ is the relative position of antenna $n$ with respect to the array center
\begin{align}
  \label{eq:antennaRelativePosition_x}
  x_n & = 0,\\
  \label{eq:antennaRelativePosition_y}
  y_n & = d \left(-\frac{\left(\sqrt{N}-1\right)}{2} + \textrm{mod} \left(n-1, \sqrt{N}\right)\right),\\
  \label{eq:antennaRelativePosition_z}
  z_n &= d\left( -\frac{\left(\sqrt{N}-1\right)}{2} + \left\lfloor\frac{n-1}{\sqrt{N}}\right\rfloor\right).
\end{align}
Equations \eqref{eq:arrayPosition_x} and \eqref{eq:antennaRelativePosition_x} state that our array is arranged on the $yz$ plane so that, strictly speaking, a 2D notation should be enough  to represent the array configuration. However, throughout the paper we adopt the $\mathbb{R}^3$ formulation for the sake of generality.

In our numerical results, we will model the element radiation pattern $g\left(\theta\right)$ in~\eqref{eq:arrayVector}, as a function of the elevation $\theta$ only, assuming a circular-symmetric \ac{DRA} \cite{Element_Model}: $g\left(\theta\right)=\sqrt{\Gamma} \cos^q(\theta)$, where $q=(\Gamma-2)/4$, and
$\Gamma=4 \pi A \mu$ is the peak gain of the radiating element, with $A$ and $\mu$ denoting the \ac{DRA} elementary cell area and the (linear) radiating element efficiency, respectively. However, our approach applies to any \ac{FoA} and any element radiation pattern -- the relevant performance can be evaluated using the Matlab code available from the authors upon request. 
\subsection{Analysis of the radiation pattern}
\label{model:pattern} 
Let us now consider the satellite downlink. Our \ac{FoA} will be used to generate multiple, adjacent beams serving a population of users, as we will detail in the next section. In the following, we will review well-known results for the convenience of the reader and to establish a simple notation, starting with the computation of the  \emph{array response} $\zeta(\varphi,\theta) = \left|\frac{1}{g(0)\sqrt{NS}}\mathbf{a}^H(0,0)\mathbf{a}(\varphi,\theta)\right|^2$ of the ``main'' beam, i.e., the normalized received power density radiated by the \ac{FoA} in the generic direction $(\varphi,\theta)$ when it is aimed at  $\varphi=0,\theta=0$. By using \eqref{eq:waveVector}-\eqref{eq:antennaRelativePosition_z}, we get
\begin{align}\label{eq:arrayPattern}
    \sum_{s=1}^{S}{e^{\imagunit 2\pi (\Omega z_s+\Psi y_s)/\lambda}}
    \sum_{n=1}^{N}{e^{\imagunit 2\pi (\Omega z_n+\Psi y_n)/\lambda}},
\end{align}
where $\Omega=\sin\theta$ and $\Psi=\cos\theta\sin\varphi$ are introduced for notation convenience. By using \eqref{eq:arrayPosition_y}-\eqref{eq:arrayPosition_z} and \eqref{eq:antennaRelativePosition_y}-\eqref{eq:antennaRelativePosition_z},
and recalling that, for any integer $M\geq 1$ and real-valued $T$
\begin{equation}\label{eq:geoseries}
e^{-\imagunit \pi (M-1) T}\cdot \sum_{m=1}^{M}{
e^{\imagunit 2\pi (m-1) T}}  =
\begin{cases} \frac{\sin\left( \pi M T\right)}{\sin\left( \pi T\right) }, &T \neq 0, \\
 M, & T = 0,
 \end{cases}
 \end{equation}
we get the \ac{FoA} array response
\begin{align}\label{eq:FoAPattern}
    \zeta(\varphi,\theta) = \Bigg|
    \frac{g(\theta)}{\sqrt{NS}} \cdot
    \underbrace{\zeta^\prime(\varphi,\theta) 
    }_\textrm{\ac{FoA} factor}
    \cdot
    \underbrace{\zeta^{\prime\prime}(\varphi,\theta)    }_\textrm{array factor}
    \Bigg|^2,
\end{align}
where
\begin{align}
      \zeta^\prime(\varphi,\theta) &= 
    \frac{
    \sin\big(\pi\sqrt{S}\Omega\Delta/\lambda\big)
    \sin\big(\pi\sqrt{S}\Psi\Delta/\lambda\big)
    }
    {    
    \sin\left(\pi\Omega\Delta/\lambda\right)
    \sin\left(\pi\Psi\Delta/\lambda\right)
    },\\
    \zeta^{\prime\prime}(\varphi,\theta) &= 
    \frac{
      \sin\big(\pi\sqrt{N}\Omega d/\lambda\big)
      \sin\big(\pi\sqrt{N}\Psi d/\lambda\big)
    }
         {    
           \sin\left(\pi\Omega d/\lambda\right)
           \sin\left(\pi\Psi d/\lambda\right)
         }.
\end{align}

In writing~\eqref{eq:FoAPattern}, we have explicitly identified the contribution of the \ac{FoA} $\zeta^{\prime}(\varphi,\theta)$ and the one provided by the single array $\zeta^{\prime\prime}(\varphi,\theta)$, showing that the overall pattern is the \emph{superposition} of two different effects. The directivity of the \ac{FoA} can be augmented by either increasing the number of arrays $S$, or by enlarging the array spacing $\Delta$, in order to achieve a sufficiently narrow main beam. On one hand, increasing $S$ has the obvious downside of increasing  the cost/complexity of the system (including the requirement to maintain synchronization across the physically separated arrays, as shown in \sectionname~\ref{SubSec:Phase_instability}). Unfortunately, increasing $\Delta$ significantly increases the grating lobes of the \ac{FoA} radiation pattern, with undesirable effects in terms of inter-beam co-channel interference, as better detailed in \figurename~\ref{fig:2d_spacing}. On the other hand, increasing $\Delta$ also relaxes the constraint in terms of formation flying (satellites more spaced apart).
If we let $\Delta=\sqrt{N}d$, the \ac{FoA} becomes equivalent to a single non-sparse array hosting $NS$ radiating elements, and \eqref{eq:FoAPattern} reduces to
\begin{align}\label{eq:singleArrayPattern}
  \zeta(\varphi,\theta) = \Bigg|
  \frac{g(\theta)\sin\big(\pi\sqrt{NS}\Omega d/\lambda\big) \sin\big(\pi\sqrt{NS}\Psi d/\lambda\big)}
       {\sqrt{NS}\sin\left(\pi\Omega d/\lambda\right) \sin\left(\pi\Psi d/\lambda\right)}
       \Bigg|^2,  
\end{align}
which corresponds to the array radiation pattern of a conventional (large) array with area $NSd^2$.

In the following sections, we investigate the impact of the system parameters on the \ac{FoA} radiation pattern, also including the effects due to instability and miscalibration of the system and to alternative structures of the satellite array. For the reader's convenience, we focus on parameters (especially, concerning the spacing across radiating elements $d$) which are suitable for an \ac{R-GEO} scenario, as better detailed in \sectionname~\ref{performance:geo}. However, the insights derived in what follows are valid in general, and applicable to the \ac{LEO} scenario as well, possibly using the parameters considered in \sectionname~\ref{performance:leo}.

\subsection{\acs{FoA} radiation pattern examples}\label{model:examples}
To understand the impact of the different parameters, we focus on a carrier frequency at S-band with $f_0=2.2\,\textrm{GHz}$ and $\lambda\simeq 13.6\,\textrm{cm}$, and we assume  $N=49$ radiating elements   per array, with $d=4.5\lambda$. This leads to an array size $L\simeq3.76\,\textrm{m}$, which is compatible with the $5.4\,\textrm{m}$-diameter Ariane 6 launcher fairing (corresponding to maximum array side length of $3.82\,\textrm{m}$). We assume an \ac{FoA} with $S=169$ satellites (arrays), and we consider the following array spacings: $\Delta=\{\sqrt{N}d, 1.25L, 2.5L, 5L\}$, where the first value corresponds to the non-sparse \ac{FoA} introduced above.

\begin{figure*}[t]
  \begin{center}
    \subfigure[$\Delta=1.25L$.]{
    {\includegraphics[width=1.4\columnwidth]{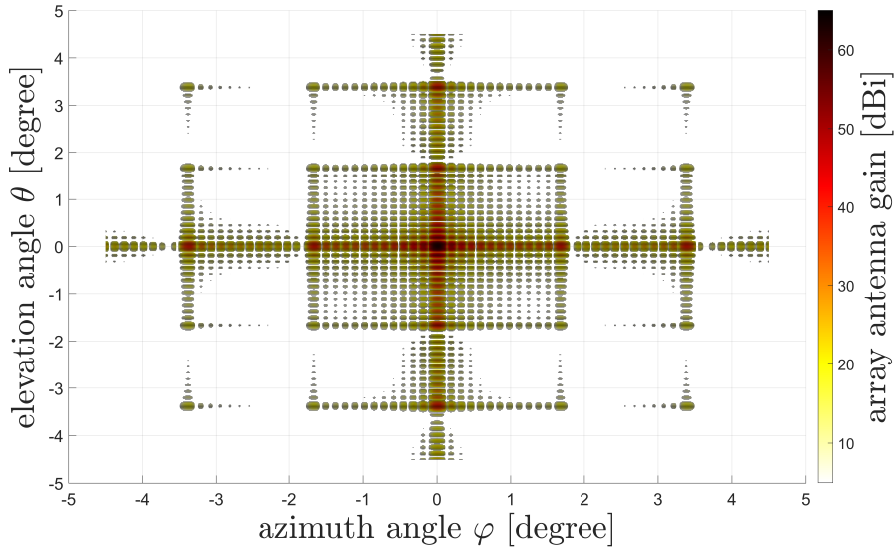}}
    \label{fig:3d_a}}
    \\
    \subfigure[$\Delta=5L$.]{
    {\includegraphics[width=1.4\columnwidth]{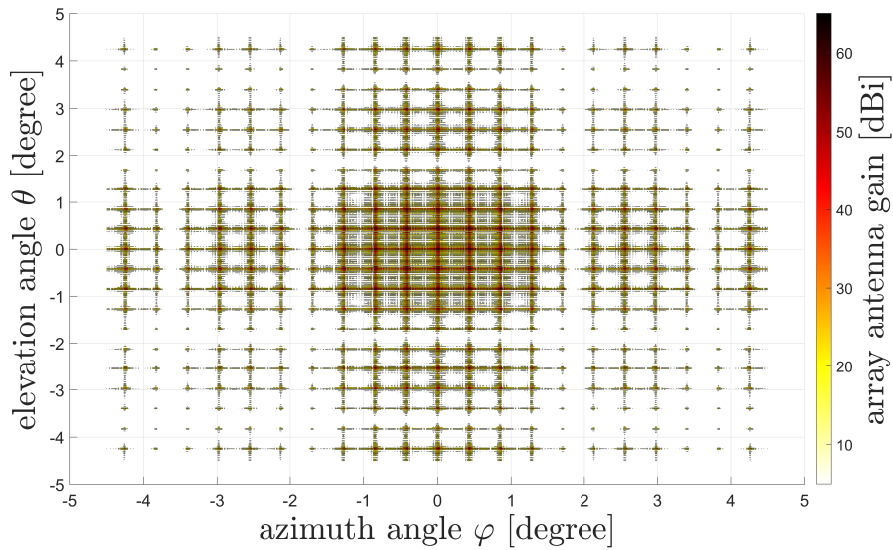}}
    \label{fig:3d_b}}
    \caption{2-D map of the \acs{FoA} gain $\zeta(\varphi,\theta)$ for different values of the array spacing $\Delta$. The range for $\theta$ and $\varphi$ is compatible with an \acs{R-GEO} scenario.}
    \label{fig:3d}
  \end{center}
\end{figure*}

\begin{figure}[t]
  \begin{center}
    {\includegraphics[width=\columnwidth]{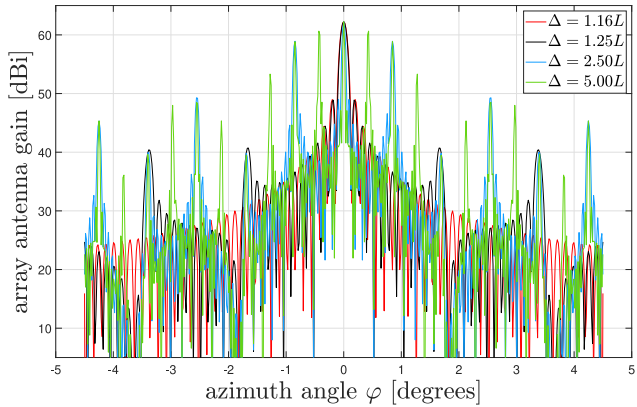}}
    \caption{\acs{FoA} gain $\zeta(\varphi,\theta=0)$ on the horizontal plane for different values of the array spacing $\Delta$.}
    \label{fig:2d_spacing}
  \end{center}
\end{figure}

\figurename~\ref{fig:3d} reports the false-color array response for the cases $\Delta=1.25L$ and $\Delta=5L$, whereas \figurename~\ref{fig:2d_spacing} depicts the array response on the horizontal plane (i.e., $\theta=0$) for different spacings -- notice the high absolute gain (around $60\,\textrm{dBi}$) of the \ac{FoA} thanks to the large number of radiating elements.

As can be seen, increasing $\Delta$ creates a very narrow main beam, but also generates very high grating lobes. The case $1.25L$ presents the best combination in terms of central beam-width and grating lobes level, and will be used as our preferred spacing in the remainder of the paper. An inter-array gap in the order of one meter, such as our $0.25L$, is feasible with the current formation flying technologies, while lower distances are probably not (see \sectionname~\ref{subsec:calibration}).

\begin{figure}[t]
  \begin{center}
    {\includegraphics[width=\columnwidth]{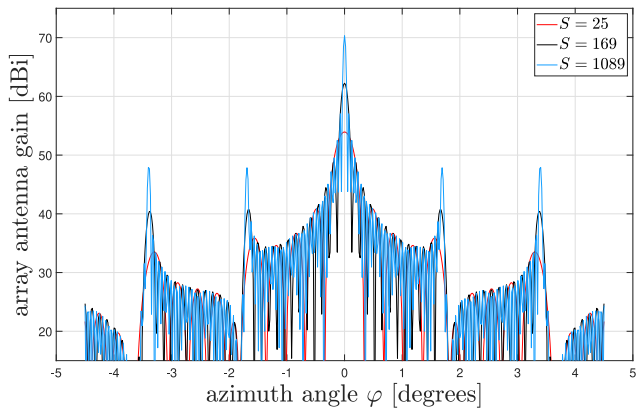}}
    \caption{\acs{FoA} gain $\zeta(\varphi,\theta=0)$ on the horizontal plane for different values of the number of satellites (arrays) $S$.}
    \label{fig:2d_arrayNumber}
  \end{center}
\end{figure}

\figurename~\ref{fig:2d_arrayNumber} shows the impact on the \ac{FoA} performance of the number of satellites $S$, using $\Delta=1.25L$. Increasing $S$ improves the performance in terms of the \ac{FoA} pattern, as the array gain increases and the central beam becomes narrower, with no increase of grating lobes. As anticipated above, increasing $S$ has a severe impact in terms of the complexity and cost of the system. The same applies of course, on a smaller scale, to the single-array parameters, $N$ and $d$. In our application examples to follow, we consider consolidated architectures for satellite arrays to determine $d$, and we consider physical constraints of the satellite launcher to determine $N$, once $d$ is given. The pattern gain is strictly related to the \ac{FoA} aperture or beamwidth, that in its turn determines the beam coverage on Earth, as we will discuss in great detail in \sectionname~\ref{sec:system}.
\subsection{Punctured \acs{FoA} using winglets}
\label{model:winglets} 
\begin{figure}[t]
  \begin{center}
    \begin{overpic}[width=\columnwidth]{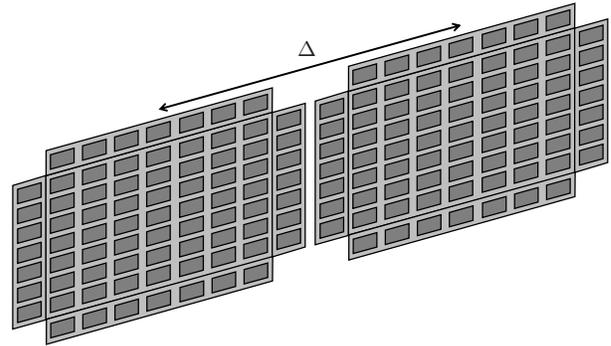}
      \put(48,47){\footnotesize{$\Delta$}}
    \end{overpic} 
    \caption{An array configuration with $N=49$ radiating elements and four winglets, each hosting $7$ radiating elements.}
    \label{fig:winglets}
  \end{center}
\end{figure}

\begin{figure}[t]
  \begin{center}
    {\includegraphics[width=0.75\columnwidth]{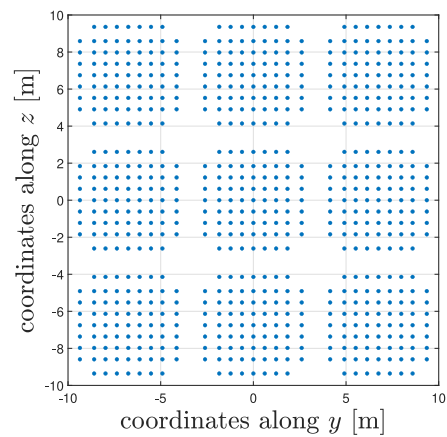}} 
    \caption{Punctured appearance of a \acs{FoA} with $S=9$ and a wingleted elementary array as in \figurename~\ref{fig:winglets}.}
    \label{fig:puncture}
  \end{center}
\end{figure}

A simple approach to extend the overall area of the \ac{FoA} for a given number of satellites $S$ is providing each satellite array with deployable ``winglets'' on each  array side -- in a sense, this is a compromise between adopting a conventional planar array and a complicated deployable antenna structure. The winglets are folded upon the main array before launch and deployed in space, so that the area occupied by each satellite with folded winglets at launch time is the same as without winglets.\footnote{Actually, the folded winglets increase the satellite height at launch time, thus occupying a slightly larger launcher fairing volume.} Once deployed in space, the array has a cross-like shape with a total area larger than that of the main array, so that the satellites in the \ac{FoA} are more spaced apart, and the total \ac{FoA} size is larger. In so doing, the \ac{FoA} is ``punctured'' by the empty area created by the cross-like shapes: an example of a winglet configuration is shown in \figurename~\ref{fig:winglets}, wherein each main array is composed by $N=49$ radiating elements and it is supplemented by four additional $7$-element winglets, for a total of $N^\prime=49+4\times7=77$ elements. Keeping the same spacing $d$ {between the elements} of the main array, the inter-array spacing $\Delta$ is larger than without winglets. {However, if we consider the wingleted-arrays, the closest distance between elements in the winglets of two adjacent arrays remains the same $\Delta$. To visualize it,} \figurename~\ref{fig:puncture} shows the punctured appearance of the \ac{FoA} with wingleted-arrays for the same base array as in \figurename~\ref{fig:winglets} and with $S=9$. This arrangement is more convenient than a standard $N=9\times9=81$ one, with a number of radiating elements similar to $77$: for the same inter-element distance $d$, the conventional $9\times9$ array requires a larger launcher fairing -- this is the very advantage of using the wingleted-array configuration, or, as an alternative, a larger array is obtained with the same fairing size. Note also that, while the winglets in the example reported in \figuresname~\ref{fig:winglets}-\ref{fig:puncture} host only one row of $\sqrt{N}=7$ radiating elements, each winglets can in general host up to $\left(\sqrt{N}-1\right)/2$ rows of $\sqrt{N}$ radiating elements, while ensuring proper unfolding after the launch. This approach is considered in the numerical results presented in \sectionname~\ref{performance:winglets}.



\begin{figure}[t]
  \begin{center}
    {\includegraphics[width=\columnwidth]{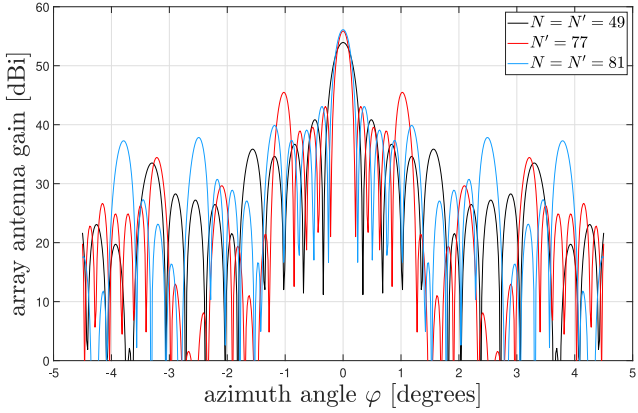}}
    \caption{\acs{FoA} gain $\zeta(\varphi,\theta=0)$ in the horizontal plane for different array configurations  ($S=25$).}
    \label{fig:wingletPattern}
  \end{center}
\end{figure}

To assess the benefits of the winglets, we can extend the analysis in \sectionname~\ref{model:pattern} to the array with winglets, by simply including the location of each winglet element into \eqref{eq:arrayVector}. The performance of the punctured array with winglets is analyzed in \figurename~\ref{fig:wingletPattern}, in which the square array configuration depicted in \figurename~\ref{fig:FoA} is considered, using again $\lambda\simeq 13.6\,\textrm{cm}$ ($L\approxeq3.68\,\textrm{m}$ with $N=49$), $S=25$, $d=4.5\lambda$, and $\Delta=1.25L$. In particular, black and blue lines report the radiating pattern for two cases with no winglets (the usual $N=N^\prime=49$ array and a larger $N=N^\prime=81$ array, respectively), whereas the red line depicts the case with $N=49$ and one row of $7$ radiating elements per winglet ($N^\prime=N+4\times7=77$ radiating elements per array). 
The wingleted-array is an intermediate case, as it shares with the usual array (black line) the same number of central radiating elements (area at launch), whereas it shares approximately the same number of overall radiating elements (therefore \ac{FoA} gain) with the larger array (blue line). The punctured \ac{FoA} has a larger (peak) gain than its conventional counterpart without winglets for the same total number of radiating elements, due to the overall area increase. It also exhibits higher grating lobes because of the larger degree of sparseness introduced by the puncturing pattern. The advantage of such configuration wrt our communication-related \acp{KPI} will be evaluated in \sectionname~\ref{sec:results}. 
\subsection{Impact of \acs{FoA} instability/miscalibration}
\label{SubSec:Phase_instability}
When actually implementing the \ac{FoA} architecture, a major difficulty arises in ensuring phase stability (calibration) across arrays, since they are hosted by different satellites with coordinated, but individual attitude control, and since they are fed by coordinated but individual \ac{RF} transmitters. It is therefore pivotal to understand the antenna pattern sensitivity to possible calibration errors among the various \ac{FoA} chains. To this aim, we introduce a random phase shift $\phi_s$ that is specific of the $s$-th satellite (array) modeling the effect of such instability. The array response therefore becomes $\zeta(\varphi,\theta) = \left|\frac{1}{g(0)\sqrt{NS}}\mathbf{a}^H(0,0)\tilde{\mathbf{a}}(\varphi,\theta,\boldsymbol{\Phi})\right|^2$ where
\begin{align}\label{eq:precoding_vector_subarray}
  \tilde{\mathbf{a}}\left(\varphi, \theta, \boldsymbol{\Phi}\right) \in {\mathbb{C}^{NS}} =
  \left[ \tilde{\mathbf{a}}_1^T\left(\varphi, \theta, \phi_1\right), \dots,
    \tilde{\mathbf{a}}_S^T\left(\varphi, \theta, \phi_S\right) \right]^T,
\end{align}
with $\tilde{\mathbf{a}}_s\left(\varphi, \theta, \phi_s\right)=e^{\imagunit \phi_s} \mathbf{a}_s\left(\varphi, \theta\right)$, and $\boldsymbol{\Phi}=[\phi_1,\dots,\phi_S]^T$.  The entries of the latter vector are a set of independent, identically distributed random variables representative of errors in the satellite spatial/electrical alignment/determination.

\begin{figure}[t]
  \begin{center}
    {\includegraphics[width=\columnwidth]{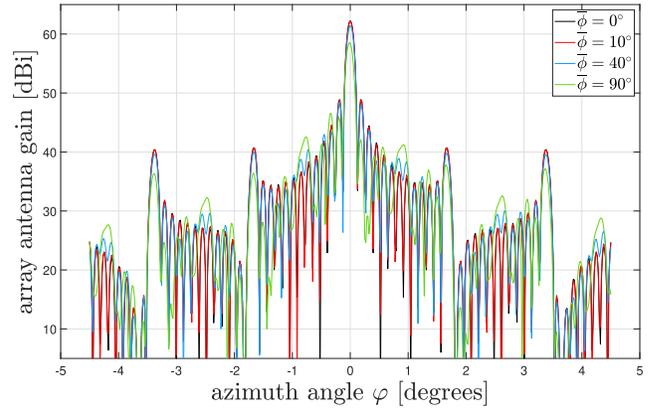}}
    \caption{2D \acs{FoA} gain $\zeta(\varphi,\theta=0)$ as a function of the maximum phase instability per array. $\lambda\simeq13.64\,\textrm{cm}$, $S=169$, $N=49$,
$d=4.5\lambda$ ($L\simeq3.68\,\textrm{m}$), and $\Delta=1.25L$).}
    \label{fig:phaseInstability}
  \end{center}
\end{figure}

\figurename~\ref{fig:phaseInstability} shows the sensitivity of the \ac{FoA} pattern gain to phase inaccuracy. Each plot is computed assuming that $\left\{\phi_s\right\}_{s=1}^{S}$ is uniformly distributed in
$\left[-\overline{\phi}, +\overline{\phi}\right]$,  $\overline{\phi}=0^\circ$ (black line) representing the case with perfect calibration. The other cases represent increasing instability with $\overline{\phi}=10^\circ$, $\overline{\phi}=40^\circ$, and $\overline{\phi}=90^\circ$ shown in red, blue, and green lines, respectively. 

To better emphasize the impact of phase instability, only \emph{one} random realization per value is considered. We performed many trials with independent realizations, and we always found similar results to the one shown here, that can be considered typical. We see that the higher $\overline{\phi}$, the larger the asymmetry of the pattern, but the impact of mis-calibration on the overall \ac{FoA} gain is not disruptive.  Furthermore, the sensitivity to random phase errors is reduced by increasing $S$.

\subsection{{\acs{FoA} feeding network}}
Before delving into performance analysis of the communication network, we have to tackle a final fundamental issue regarding how the \ac{FoA} receives/sends out the very many signals that are either transmitted in the downlink or received in the uplink. This is an issue indeed since the \ac{FoA} technology is intended to handle a very large throughput/bandwidth, so that managing such aggregated wideband signal is not trivial. Just to make and example, assume that each satellite handles a chunk of $60$ MHz bandwidth at S-band, and that our \ac{FoA} support few thousand of beams -- we end up with an aggregated bandwidth of several hundreds of GHz to feed the \ac{FoA} from Earth. The problem is similar to what designers face today when trying to deploy and operate a \ac{VHTS} \cite{VHTSref}  whose total handled throughput is designed to be in the order of several hundreds of Gbps. In addition to Ka-band feeds, such demanding configuration may also need free-space optical links to maximize throughput while keeping the number of gateways limited to an acceptable amount.

We will quantify this requirement and propose a solution later on in \sectionname~\ref{feednet} (based on a feeding, central satellite that drives via \acp{ISL} all of the smaller satellites in the formation), when we present study cases with specific figures for the \ac{FoA} configuration and the bandwidth to be handled.

\section{Communications network modeling and \acsp{KPI}}
\label{sec:system}
The \ac{FoA} technology can be used, together with appropriate beamforming, to create a tight multi-beam footprint on Earth to provide digital connectivity to a population of mobile users, equipped with a 4G-like handheld mobile terminal. The goal is to come as close as possible to the requirement in terms of throughput/unit area $\rho$ (measured in Mbps/km$^2$) of a terrestrial 4G cellular network serving a rural area with modest user density. We limit our ambition to this, as the
preliminary results reported in \sectionname~\ref{Sec:Introduction_and_Motivation} show that truly 5G broadband mobile figures are not achievable with low-cost, low-weight hand-held terminals and acceptable size \ac{FoA}.

In the following we adopt a plain beamforming to generate a regular beam pattern without any attempt to adaptively minimize the other beam sidelobes effects by means of pre-coding techniques. This assumption is justified by the need to keep payload complexity within affordable limits and the potential limited performance gain achievable in practice. In fact, some recent work (e.g., \cite{Angeletti_P-MMIMO,Angeletti_H-RRM}) shows that the advantage of ideal \ac{MMSE} pre-coding in a realistic satellite multi-beam scenario is rather limited compared to optimized high overlapped regular beam pattern with high-performance linear complexity \ac{RRM} \cite{Angeletti_H-RRM}. This pragmatic \ac{mMIMO} approach has sizeable complexity and implementation advantages, in particular when the traffic scenario is highly dynamic, as in the \ac{LEO} case.
\subsection{Network configuration}
\label{system:configuration}
We consider a multi-beam satellite network with frequency reuse to provide spatially continuous service provisioning over the covered area. The maximum coverage area is determined by the satellite (one-sided) coverage angle $\overline{\theta}$ given by \cite{SatSystems2010}
\begin{align}\label{eq:communicationAngle}
  \overline{\theta} =  \sin^{-1} \left( \frac{R_\mathsf{e}}{R_\mathsf{e}+h_\mathsf{sat}} \cos\epsilon \right),
\end{align}
where $R_\mathsf{e}=6,371\,\textrm{km}$ is the Earth radius, $h_\mathsf{sat}$ is the satellite height above the ground, and $\epsilon$ is the minimum elevation angle for the user terminal on Earth to receive the satellite signal. The value  $\overline{\theta}$ will represent our investigation boundary for the \ac{FoA} and network performance. 
Thanks to the symmetry of the \ac{FoA}, coverage will be the same in the azimuth direction: $\overline{\varphi}=\overline{\theta}$. Using the values introduced in \tablename~\ref{tab:Example_GEO_LEO_sizing} for the satellite orbits, we get $\overline{\theta}\approxeq4.3^\circ$ for the regional \ac{GEO} and  $\overline{\theta}\approxeq44.8^\circ$ for \ac{LEO}. The coverage area is assumed to be circular, with radius $h_\mathsf{sat}\tan(\overline{\theta})$.

Each beam serves a certain number of \acp{UT}, with an unspecified multiplexing technology based on  orthogonal resources in  time, in frequency, and/or in both. Sticking to a  conventional frequency-reuse  coloring scheme with reuse factor $M$ \cite{rappaportwireless}, all beams are continuously transmitting with a reduced bandwidth allocation, so that the aggregate bitrate per beam is correspondingly reduced by the number $M$ of frequency sub-bands that are adopted. 

We will also assume to operate the downlink in the S-band using a total available bandwidth $B$ \cite{EC_S-band_LEX}. This frequency band, next to the terrestrial mobile networks frequency allocation, facilitates the reuse of the terrestrial \ac{UT} antenna and \ac{RF} front-end. The disadvantage is the relatively limited available bandwidth $B$ (a few tens of MHz) that forces to keep the reuse factor $M$ as low as possible so that the beam/\ac{UT} peak bitrate is not reduced too much. For this reason, we adopt in the following either $M=1$ or $M=3$, thus limiting the potential benefits deriving by possible dynamic \ac{RRM}.
\subsection{Channel model, payload and beamforming}
\label{system:channel}
Based on the network model above, we partition our service area into a maximum number $\overline{K}$ of beams, that depends on the coverage area radius and the reuse factor $M$, with beam $k$, $k=1,\ldots,K\le\overline{K}$ serving a certain number of \acp{UT}. In the following we will analyze inter-beam interference in the user downlink, concentrating on the sub-population of all of the users that are allocated to the same orthogonal resource in each beam (those who can reciprocally create inter-beam interference). Within this sub-population, the user identifier clearly coincides with the beam identifier $k$.

We assume a Ku- or Ka-band feeder link from the gateway Earth station to the satellite (or from another satellite in the constellation when the \ac{LEO} satellite is not in contact with the ground) with high-gain antenna, so that the feeder link \ac{SNR} is not significantly affecting the overall forward-link performance. We also consider a user downlink bandwidth $B=60\,\textrm{MHz}$ (neglecting possible flux density limitations deriving from applicable regulations) as a 5G-oriented assumption. More details about system parameters can be found in \sectionname~\ref{sec:results}. 


The \ac{UT} is equipped with a low-gain omni-directional antenna, similar to the ones used for terrestrial hand-held phones. The corresponding $G/T=-31.6\,\textrm{dB/K}$ is in line with 3GPP's \ac{NTN} 5G current assumptions \cite{3GPP_38221}. In addition, a $1$-dB demodulator implementation loss is also considered.

We assume an ideal payload composed of an on-board processor implementing the appropriate beamforming which routes all of the active \acp{UT}' resource elements to the appropriate beams. We denote by $\mathcal K \subseteq \{1,2,...,\overline{K}\}$ the set of active beams, $|\mathcal K| = K \le \overline{K}$, and $s_k $ the \ac{DL} data signal intended for the \ac{UT}~$k\in \mathcal K$, and $p_k$ the $k$-th transmitted power. We call $\left(\varphi_k^c, \theta_k^c\right)$ the azimuth and elevation angles of the $k$-th active beam that is typically directed towards the center of the coverage area. The signal $s_k$ intended for the UT $k\in\mathcal{K}$ is subject to a beamforming vector $ {\bf v}_k \in \mathbb{C}^{NS}$ that routes $s_k$ towards its appropriate beam $k$:
\begin{equation}
\label{eq:precodingVector}
{\bf v}_k = \frac{{\bf a}(\varphi^c_k, \theta^c_k)}{\sqrt{\alpha}},
\end{equation}
where $\alpha$ is a normalization factor adopted by the \ac{FoA}. In particular, throughout the paper we consider $\alpha=\textrm{tr}\left({\bf V}^H{\bf V}\right)$, where ${\bf V}=\left[ {\bf v}_1, \dots, {\bf v}_K  \right]$ is the matrix collecting all beamforming vectors.

The received signal by the \ac{UT} $k$ is modeled as 
\begin{equation}
  y_k = \sqrt{\beta_k}{\bf a}^H(\varphi_k, \theta_k){\bf x} + n_k
\end{equation}
where $\beta_k$ is the attenuation experienced by the transmitted signal, $(\varphi_k, \theta_k)$ are the azimuth and elevation angles of \ac{UT} $k$, $n_{k} \sim \CN(0,\sigma^2)$ is \ac{AWGN} with variance $\sigma^2$, and ${\bf x} \in \mathbb{C}^{NS}$ is the \ac{DL} signal transmitted by the \ac{FoA}
\begin{equation}
  {\bf x} = \sum_{k \in \mathcal K} {\bf v}_ks_k.
\end{equation}
According to our model, the \ac{SINR} experienced by \ac{UT} $k$ is
\begin{equation} \label{eq:sir}
  \gamma_k = \frac{ |{{\bf a}^H(\varphi_k, \theta_k)}{{\bf v}_k}|^2 p_k}{\sigma^2/\beta_k+\sum_{i \in \mathcal K, i\ne k}{|{{\bf a}^H(\varphi_k, \theta_k)}{{\bf v}_i}|^2 p_i}},
\end{equation}
which will be used as a key factor to compute the communication-related performance in \sectionname~\ref{sec:results}. The different parameters in the \ac{SINR} computation (\{$\beta_k$\},\{$p_k$\},$\sigma^2$) result from system assumptions and a detailed link budget computation, as described in the next section.
\subsection{User traffic model}
\label{system:traffic}
\begin{figure}[t!]
  \begin{center}
    \subfigure[T1 model.]{
    {\includegraphics[width=0.6\columnwidth]{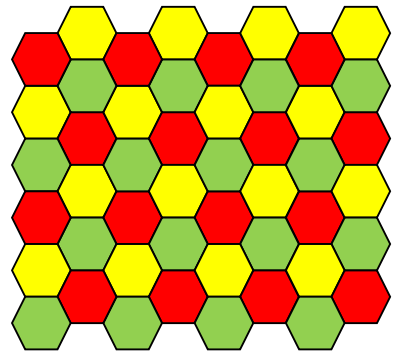}}
    \label{fig:cells_T1}}
    \\
    \subfigure[T2 model.]{
    {\includegraphics[width=0.72\columnwidth]{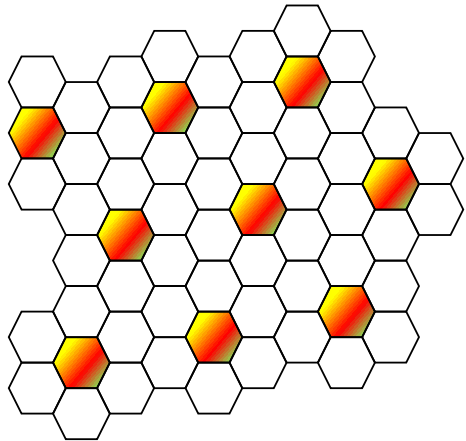}}
    \label{fig:cells_T2}}
    \caption{Beam placement and user allocation as functions of the user traffic model.}
    \label{fig:cells}
  \end{center}
\end{figure}

In our overall efficiency computation, we will consider two different scenarios for the user-generated data traffic and for our system layout:
\begin{itemize}

\item \emph{Uniform, continuous traffic across the whole satellite coverage area (T1)}: In this case, we instantiate a large number of active contiguous beams with a relatively small (frequency or time) reuse factor $M=3$ (see \figurename~\ref{fig:cells_T1}) -- a universal frequency reuse with $M=1$ is considered to be too ``aggressive'' as it leads to a problematic link budget event with low-rate error protection coding and small-cardinality signal constellations. The $M=3$ different colors in \figurename~\ref{fig:cells_T1} correspond to the different subsets of orthogonal resources (sub-bands).

\item \emph{Isolated, bursty spots of traffic (not requiring continuous coverage) (T2)}: In this case we assume for simplicity the presence of  regularly spaced uniform traffic ``islands'' matching the beam size. In particular, we assume that each active beam is at the center of a cluster of $C=7$ beams, with the surrounding $6$ being inactive (see \figurename~\ref{fig:cells_T2}). In this case, the reuse factor may be $M=1$, allocating all the available bandwidth to the active/(central) beam. The active beams in \figurename~\ref{fig:cells_T2} are all-color, meaning that all resources are available and re-used in each beam. We are aware that this regular traffic cluster geometry may not be very realistic, as actual  clusters will be irregularly distributed. However, this model  is considered appropriate enough for our  scenario assessment. 
\end{itemize}

\subsection{\Acl{KPI}: network throughput}
\label{system:throughput}
Evaluation of the \ac{SINR} as in \eqref{eq:sir} according to the traffic models of scenarios T1 and T2 is the key to deriving our main \ac{KPI}, i.e., the network throughput $\rho$ as outlined in the sequel.

Assuming Gaussian statistics for the inter-beam interference, the spectral efficiency of the user link can be evaluated according to 3GPP's 5G \ac{NTN} performance specification \cite{3GPP_38221} in \tablename~\ref{tab:5G_NTN_PHY_parameters}, which maps the user \ac{SINR}  to a spectral efficiency $\eta$. The network throughput in terms of total bitrate per beam will just be the product between this spectral efficiency $\eta$ and the assumed bandwidth $B/M$ in a beam ($M$ the reuse factor), i.e., $B \eta /M$. 
\begin{table*}[h]
    \centering
    \small{
    \begin{tabular}{c | c c c c c c c c c c c}\hline
        \emph{Required \ac{SINR} at} & 
        \multirow{2}{*}{-7} &
        \multirow{2}{*}{-6} &
        \multirow{2}{*}{-5} &
        \multirow{2}{*}{-4} &
        \multirow{2}{*}{-3} &
        \multirow{2}{*}{-2} &
        \multirow{2}{*}{-1} &
        \multirow{2}{*}{0} &
        \multirow{2}{*}{1} &
        \multirow{2}{*}{2} &
        \multirow{2}{*}{3} \\
        \emph{\acs{BLER}=$10^{-2}$ [dB]}     &&&&&&&&&&&\\\hline
        \emph{Spectral efficiency $\eta$} &
        \multirow{2}{*}{0.20} &
        \multirow{2}{*}{0.23} &
        \multirow{2}{*}{0.31} &
        \multirow{2}{*}{0.38} &
        \multirow{2}{*}{0.49} &
        \multirow{2}{*}{0.59} &
        \multirow{2}{*}{0.69} &
        \multirow{2}{*}{0.82} &
        \multirow{2}{*}{0.95} &
        \multirow{2}{*}{1.11} &
        \multirow{2}{*}{1.28} \\
        \emph{[b/s/Hz]}                       &&&&&&&&&&&    \\\hline
    \end{tabular}
    }
    \caption{3GPP 5G \acs{NTN} \acs{PHY} performance on the \acs{AWGN} channel.}
    \label{tab:5G_NTN_PHY_parameters}
\end{table*}

The aggregate network throughput across the $K$ active beams will be $K \cdot B \eta /M$, and the  final datum of bitrate per km$^2$ $\rho$ will be the aggregate throughput divided by the coverage area (depending in its turn on the coverage angle of the \ac{FoA}). This computation is a bit complicated by the necessity to set a reasonable number of satellites forming the \ac{FoA}, and to take into account all of our assumptions or constraints about available power, channel status, etc. Therefore the \ac{KPI} $\rho$ is found following a series of steps summarized as follows:
\begin{enumerate}

\item A minimum target bitrate/beam over a certain coverage area is set as a worst-case design goal. 

\item The key system parameters corresponding to the scenario of interest are established: orbit, user link frequency, assigned bandwidth,  satellite platform, payload, array size, traffic model, etc...) - we will detail such parameters in the next Section.

\item The number of satellites $S$ for the \ac{FoA} is fixed, so that the \ac{FoA} gain and coverage angle is determined. A reasonable  value is that leading to a beam diameter comparable to the size of a terrestrial rural cell.

\item The available DC power  for each satellite is derived.

\item The available RF power for each satellite array is computed.

\item  An initial number of active beams $K=\overline{K}$ is computed, according to the beam size as above and to the extension of the target coverage area. 

\item The \ac{PHY} configuration from \tablename~\ref{tab:5G_NTN_PHY_parameters} that allows to attain the minimum target bitrate per beam for the most critical link budget configuration is selected. The same configuration is retained also for the less critical  cases. \label{lbudg}

\item The link budget, including the interference coming from the other active beams, is computed, leading to the evaluation of the \ac{SINR} of the user, whose location is randomly selected within the beam of interest. Either one of the following cases can occur: \label{sinreval}

\begin{enumerate}  

\item The link margin is positive. In this case we get back to step \ref{lbudg}  selecting the next, more spectral efficient \ac{PHY} configuration. 
    
\item The link margin is negative. Then, the number of beams is reduced starting from the outermost ones (the power/beam is increased but the  coverage area is correspondingly reduced) proportionally to the inverse of the current link margin (in linear units) and step \ref{sinreval} is repeated until the margin is positive again, but so small that a more efficient PHY configuration cannot be selected. 

\end{enumerate}

\item 
At the end of this iterative procedure, the number of active beams $K$ and the network throughput $\rho$ {are obtained} under the relevant system assumptions and constraints. With these values, the bitrate/beam and the (both absolute and normalized per area) network throughput are computed.

\end{enumerate}

In the following section, we will present detailed results derived after this procedure in the two reference cases (\ac{GEO}/\ac{LEO}) that we have already introduced.

\section{\acs{FoA}/network configuration and performance results}
\label{sec:results}
\subsection{Description of use cases}\label{results:usecases}
As already mentioned, we will focus our attention on two main use cases: a \ac{GEO} satellite intended for regional service (\ac{R-GEO}) and a \ac{LEO} satellite in a global-coverage constellation.  As already mentioned, our target is to design the \ac{FoA} in order to provide an acceptable performance to a standard hand-held terminal. 

The array spacing is selected to avoid grating lobes in the coverage area, introduced in \sectionname~\ref{system:configuration}. The reason for the different array elements spacing in these two study cases is related to the larger \ac{LEO} array field of view compared to the \ac{R-GEO} case. The approximate equation to compute the normalized element spacing $d/\lambda$ is \cite{Mailloux}
\begin{align}
\frac{d}{\lambda} \le \frac{1}{2\sin\left(\overline{\theta}\right)},
\label{eqn:array_spacing}
\end{align}
where $\overline{\theta}$ represents the coverage angle \eqref{eq:communicationAngle} of the \ac{FoA}.

This also impacts, together with all the parameters introduced in \sectionname~\ref{sec:model} (notably, the number of radiating elements per satellite $N$ and the number of satellites $S$), the array response vector $\mathbf{a}(\varphi,\theta)$ \eqref{eq:FoAVector}, that is used to numerically compute the on-ground beam radius $R$. In particular
\begin{align}\label{eq:beamRadius}
  R=h_\mathsf{sat}\tan(\varphi_{-3\,\textrm{dB}}),
\end{align}
where $h_\mathsf{sat}$ is the satellite height introduced in \sectionname~\ref{system:configuration}, and $\varphi_{-3\,\textrm{dB}}$ is the azimuth angle such that $\|\mathbf{a}(\varphi_{-3\,\textrm{dB}},0)\|^2/\|\mathbf{a}(0,0)\|^2=1/2$.

The size of the \ac{FoA} is also crucial for the determination of the available onboard power. In particular, we assume that each  array has a flat shape with solar cells mounted on the reverse side of the array~pointing to the Earth, so that the array area determines the available \ac{DC} power. For simplicity, we optimistically assume that the satellite power subsystem is able to generate a certain average \ac{DC} power for each \ac{SG} square meter, irrespective of the variation of the attitude and position of the satellite

We also make realistic assumptions about the split of the available satellite DC power among the different subsystems (e.g. platform, digital processor, receiver front-end) to evaluate the available  transmit \ac{RF} power. In both use cases (\ac{LEO}/\ac{GEO}) we assume to operate the \ac{SSPA} at $1$-dB compression point with an \ac{SSPA} efficiency of about $40\%$. For this reason, the overall \ac{DC}-to-\ac{RF} conversion efficiency efficiency is equal to $(40-5)\%=35\%$. Further assumptions on the satellite platform power  are summarized in \tablename~\ref{tab:Platform_power_assumptions}.
\begin{table}[t]
    \centering
    \small{
    \begin{tabular}{l c c c}\hline
      \emph{Parameter}                & \emph{\acs{R-GEO}} & \emph{\acs{LEO}} & \emph{Unit}  \\ \hline
      Solar \acs{DC} power/unit surface &  $200$               & $200$              & W/m$^2$\\
      Array surface                     &  $13.56$             & $14.16$            & m$^2$  \\
      Solar \acs{DC} power/array        &  $2711$              & $2833$             & W      \\
      Tx platform \acs{DC} power ratio  &  $40$                & $25$               & \%     \\
      Max platform \acs{DC} power per
      Tx array                          &  $1084$              & $708.25$           & W     \\
      Tx \acs{DC}-to-\acs{RF} conversion
      efficiency                        &  $35$                & $35$               & \%     \\
      Max Tx \acs{RF} power available
      per array                         &  $379.6$             & $247.9$            & W      \\\hline
    \end{tabular}
    }
    \caption{Satellite platform assumptions.}
    \label{tab:Platform_power_assumptions}
\end{table}

To make our analysis as realistic as possible, on top of the downlink free-space path loss $\beta_k$, we applied  additional signal attenuation terms. In particular, we  included an (average) $0.5$-dB atmospheric loss, a user body adsorption loss of $3\,\textrm{dB}$, and a fading margin of $3\,\textrm{dB}$ on top of the ideal \ac{AWGN} channel link budget to account for the effect of mild, slow Ricean fading effects. Then, we made a distinction between two  different channel conditions that we label C1 and C2:
\begin{itemize}

\item {\em C1: tree shadowing}: 
We add in this case a further (average) $4$-dB tree shadowing loss. 

\item {\em C2: clear path (no shadowing)}: 
This case corresponds to a line-of-sight link with no further attenuation in addition to the ones above.

\end{itemize}
Both channel conditions affect the useful as well as the co-channel interfering signal(s) coming from the other beams. 

\subsection{Performance results: Evaluation of the \ac{SINR} and preliminary throughput computation}
\begin{figure}[t]
  \begin{center}
    {\includegraphics[width=\columnwidth]{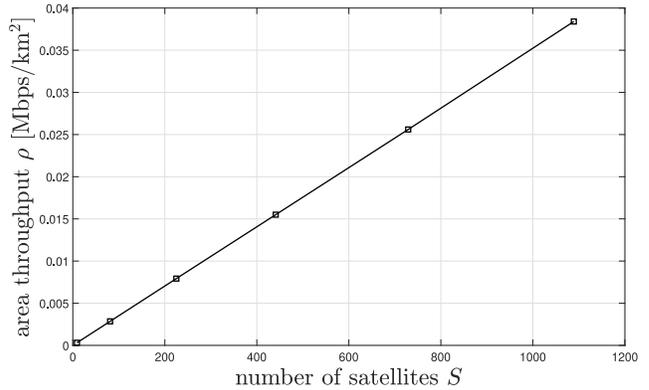}}
    \caption{Achievable area throughput $\rho$ as a function of the number of satellites $S$ (\acs{R-GEO} scenario).}
    \label{fig:rhoVsS}
  \end{center}
\end{figure}

To understand the capability of our analysis and its output, we do not set at first any throughput specification to be met, but we just analyze the resulting system efficiency in terms of bitrate/unit area $\rho$ as a function of the number of satellites in the \ac{FoA}. \figurename~\ref{fig:rhoVsS} shows the achievable $\rho$ in a simplified test-and-validation \ac{GEO} case with $7\times7$ arrays and  with the usual power assumptions, but with no mobile channel impairments (i.e. AWGN channel with no shadowing) and in traffic scenario T1.
Below the minimum number of satellites $S=9$ reported in the plot, the link budget cannot be closed and no data are available. Beyond this threshold, the efficiency $\rho$ just increases as expected with the number of satellites, coming close to values similar to those of a terrestrial cellular system ($\rho \simeq 0.1$ Mbps/km$^2$). It should be remarked that increasing the number of satellites composing the \ac{FoA}, means reducing the beam size and increasing the number of active beams to cover the same area on ground. {This result shows the scalability allowed by the \acp{FoA} concept, i.e., the number of satellites can be increased progressively to match the growing traffic demand.\footnote{This \ac{FoA} size scaling approach will have an impact on the minimum data rate supported in the return link whose analysis is outside the scope of the paper.}}

\subsection{Performance results: \Acl{R-GEO}}\label{performance:geo}
We tackle now the realistic case of a regional (continental) \ac{GEO} \ac{FoA} serving hand-held mobile terminals operating at S-band. The \ac{FoA} parameters considered for this example are summarized in \tablename~\ref{tab:R-GEO_antenna_parameters}. We compute the maximum array element spacing by using $\overline{\theta}=4.3^{\circ}$ in \eqref{eqn:array_spacing}, thus getting $d/\lambda\le 6.7$. To avoid grating lobes occurring too close to the border of the coverage area, we select $d/\lambda =4.5$, which, using an array length $L=3.76\,\textrm{m}$ (based on the considerations of the fairing size in \sectionname~\ref{model:examples}), yields $N=49$.

The orbital height is $35,870$ km and the satellite elevation is $60^{\circ}$ corresponding to a free space loss of $190.8$ dB. As previously mentioned, the array size for each satellite is determined by the accommodation space in the launcher fairing. 
\begin{table}[t]
    \centering
    \small{
    \begin{tabular}{l c c c}\hline
        \emph{Parameter}                   & \emph{Symbol}          & \emph{Value} & \emph{Unit} \\\hline
        Output back-off                      &                          & $2.0$          & dB    \\
        Antenna losses                       &                          & $1.3$          & dB    \\
        Array element gain                   &  $\Gamma$                & $23$           & dBi   \\
        Array element spacing                &  $d/\lambda$             & $4.5$          &       \\
        Number of array elements             &  $N$                     & $49$           &       \\
        Array length                         &  $L$                     & $3.68$         &  m    \\
        Array area                           &                          & $13.56$        &  m$^2$\\
        Array to array gap                   &                          & $0.92$         &  m    \\
        Number of satellites                 &  $S$                     & $1,089$        &       \\
        Total \acs{FoA} area                 &                          & $14,762$       &  m$^2$\\
        \acs{FoA} overall antenna directivity&                          & $66.3$         & dBi   \\
        \acs{FoA} maximum scanning angle     &  $\overline{\theta}$     & $\pm 4.3$      & degrees\\\hline
    \end{tabular}
    }
    \caption{\acs{FoA} parameters for the regional \acs{GEO} case.}
    \label{tab:R-GEO_antenna_parameters}
\end{table}

Following the approach that we have described in the previous section, we derived the results shown in  \tablename~\ref{tab:R-GEO_System_parameters}. We see that the aggregate  throughput largely depends on the traffic and channel scenario and ranges from 73 to 328 Gbps - \textit{de facto}, the link budget is limited by the low $G/T$ ratio of the hand-held UT. The highest throughput is achieved for case T2/C2 which takes advantage of the limited co-channel interference and good channel condition with no shadowing losses. In reality, the C1 case provides a more realistic throughput estimate considering typical mobile users' operating conditions.

\begin{table*}[t]
    \centering
    \small{
    \begin{tabular}{l c c c c c}\hline
      \emph{Parameter}                            & \emph{T1, C1} & \emph{T1, C2} & \emph{T2, C1} & \emph{T2, C2} & \emph{Unit}  \\\hline
      Single satellite total output \acs{RF} power  & $239.5$         & $239.5$         & $239.5$         & $239.5$         & W     \\
      \ac{FoA} overall output \ac{RF} power         & $53.3$          & $53.3$          & $53.3$          & $53.3$          & dBW   \\
      \ac{FoA} overall \ac{EIRP}/beam               & $83.0$          & $80.4$          & $84.0$          & $84.0$          & dBW   \\
      Line-of-sight downlink \ac{SNR}               & $11.6$          & $8.9$           & $12.6$          & $12.6$          & dB    \\
      Downlink \ac{SIR}                             & $6.6$           & $6.1$           & $12.9$          & $12.8$          & dB    \\
      Total \ac{SINR}  with shadowing               & $2.0$           & $2.5$           & $4.1$           & $6.5$           & dB    \\
      \acs{PHY} spectral efficiency                 & $0.59$          & $0.59$          & $0.82$          & $1.11$          & b/s/Hz\\
      Required \acs{PHY} \ac{SINR}                  & $-2.0$          & $-2.0$          & $0.0$           & $+2.0$          & dB    \\
      Approximate beam radius on ground             & $14.2$          & $14.2$          & $14.2$          & $14.2$          & km    \\
      Single-beam throughput                        & $10.7$          & $10.7$          & $44.7$          & $60.5$          & Mbps  \\
      Number of active beams                        & $6,777$         & $12,479$        & $5,417$         & $5,417$         &       \\
      Aggregate throughput (active beams)           & $72.70$         & $133.87$        & $242.29$        & $327.98$        & Gbps  \\
      Coverage area                                 & $3,563,449$     & $6,561,647$     & $2,848,341$     & $2,848,341$     & km$^2$\\
      Area throughput $\rho$                        & $2.04$E$-2$     & $2.04$E$-2$     & $8.51$E$-2$     & $1.15$E$-1$     & Mbps/km$^2$\\ \hline
    \end{tabular}
    }
    \caption{Link budget and throughput estimation for the regional \ac{GEO} case.}
    \label{tab:R-GEO_System_parameters}
\end{table*}

The resulting values of $\rho$ are quite remarkable, thanks to a very large \ac{FoA} equivalent size (about $110\,\textrm{m}\times 110\,\textrm{m}$) obtained through the use of $1,089$ small satellites (each with an antenna aperture of about $4\,\textrm{m}$) and a large number of active beams ($5,417$-$12,479$). Depending on the scenario, each beam provides from $10$ to $60$ Mbps, to be shared by all users served by that beam, and has an on-ground beam radius $R$ \eqref{eq:beamRadius} of about $14\,\textrm{km}$, yielding a throughput $\rho$ in the range of $2.05\times10^{-2}\div1.15\times10^{-1}$, which is fully comparable to that of a terrestrial 4G rural macro-cell. This further demonstrates the challenge in getting even 4G-like services by satellite to hand-held terminals that was illustrated in \cite{DeGaudenzi5G2022}. Finally, the coverage area depends on the number of active beams for each scenario, and ranges from about $2.8$ to $6.6$ millions of km$^2$, i.e., a substantial part of a continent (the whole Europe measures about $10.2$ millions km$^2$).
\subsection{Performance results: \ac{LEO}}\label{performance:leo}
As a second system scenario, we consider a \ac{LEO} \ac{FoA} serving hand-held mobile terminals operating at S-band. In the previous \ac{GEO} case, we assumed a single very large \ac{FoA} serving a relatively large (continent-wide) region; in this \ac{LEO} case we will have as many \acp{FoA} as satellites in the \ac{LEO} (mega-)constellation serving the whole Earth.  The \ac{FoA} parameters considered for this example are summarized in \tablename~\ref{tab:LEO_antenna_parameters}. As for the \ac{R-GEO} case, we derive the array element spacing by using \eqref{eqn:array_spacing} with $\overline{\theta}=44.8^{\circ}$. In this case we get $d/\lambda\le 0.7$, and we select $d/\lambda =0.6$ to avoid grating lobes too close to the coverage area. Considering an array length $L=3.76\,\textrm{m}$ (based on the considerations of the fairing size in \sectionname~\ref{model:examples}), we get $N=2,209$ using $d/\lambda=0.6$.

To determine the number of satellites $S$, we consider a beam size comparable with a terrestrial scenario. Based on numerical calculations, using $h_\textrm{sat}=550\,\textrm{km}$ and $S=5$, using \eqref{eq:beamRadius} we get $R=3.1\,\textrm{km}$ at nadir. Even considering the elliptical shape given by the \ac{LEO} when scanning the coverage area, we get the largest semi-axis at $\overline{\theta}$ equal to $R/\cos\left(\overline{\theta}\right)=4.4\,\textrm{km}$, which is still lower than the one provided by the \ac{R-GEO} ($14.2\,\textrm{km}$). For this reason, we focus on an \ac{FoA} with $S=5$ satellites (thus somewhat violating the hypothesis of a square \ac{FoA} illustrated in \sectionname~\ref{sec:model}), composed by a central satellite contoured by $4$ satellites on each side, since larger \ac{FoA} sizes (e.g., $S=9$) yield too small beam sizes.

\begin{table}[t]
    \centering
    \small{
    \begin{tabular}{l c c c}\hline
      \emph{Parameter}                   & \emph{Symbol}   & \emph{Value} & \emph{Unit} \\\hline
      Array element gain                   &                   & $5.55$         & dBi   \\
      Array element spacing                & $d/\lambda$       & $0.6$          &       \\
      Number of array elements             & $N$               & $2,209$        &       \\
      Array length                         & $L$               & $3.76$         &  m    \\
      Array area                           &                   & $14.16$        &  m$^2$\\
      Array to array gap                   &                   & $0.94$         &  m    \\
      Number of satellites                 & $S$               & $5$            &       \\
      Total \ac{FoA} area                  &                   & $70.8$         &  m$^2$\\
      \ac{FoA} maximum scanning angle      & $\overline\theta$ & $\pm44.8$      &  degrees \\ \hline
    \end{tabular}
    }
    \caption{\acs{FoA} parameters for the \ac{LEO} case.}
    \label{tab:LEO_antenna_parameters}
\end{table}
The \ac{LEO} relevant results are presented in \tablename~\ref{tab:LEO_System_parameters}. We see that the  required \ac{LEO} \ac{FoA} array aperture is considerably smaller than the regional \ac{GEO} one -- something to be largely expected. The \ac{FoA} equivalent aperture is $71$ m$^2$ obtained by $5$ satellites, and the aggregate throughput ranges from $0.6$ to $19.5$ Gbps depending on the traffic and channel assumptions (with the more realistic C1 case the value is ranging from $579$ to $772$ Mbps). Each beam provides about $10$ Mbps to be shared by all users across a radius $R$ of about $3$ km (much smaller of course than for the \ac{GEO} case), and the area throughput $\rho=0.43$ Mbps/km$^2$ is fully comparable to a terrestrial 4G rural macro-cell's. The coverage area of the \ac{LEO} \ac{FoA} is $1,342$-$44,351$ km$^2$, so that we estimate a number of a few thousands of satellites to provide continent-wide or global coverage.

\begin{table*}[h]
    \centering
    \small{
    \begin{tabular}{l c c c c c}\hline
      \emph{Parameter}                            & \emph{T1, C1} & \emph{T1, C2} & \emph{T2, C1} & \emph{T2, C2} & \emph{Unit}  \\\hline
      Single satellite total output \acs{RF} power  & $156.4$         & $156.4$         & $156.4$         & $156.4$         & W     \\
      \ac{FoA} overall output \ac{RF} power         & $28.0$          & $28.0$          & $28.0$          & $28.0$          & dBW   \\
      \ac{FoA} overall \ac{EIRP}/beam               & $54.4$          & $53.2$          & $43.22$         & $39.2$          & dBW   \\
      Line-of-sight downlink \ac{SNR}               & $16.0$          & $14.8$          & $4.8$           & $0.9$           & dB    \\
      Downlink \ac{SIR}                             & $3.4$           & $2.8$           & $5.1$           & $5.0$           & dB    \\
      Total \ac{SINR} with shadowing                & $2.0$           & $1.9$           & $-3.0$          & $-3.0$          & dB    \\
      \acs{PHY} spectral efficiency                 & $0.59$          & $0.59$          & $0.2$           & $0.2$           & b/s/Hz\\
      Required acs{PHY} \ac{SINR}                   & $-2.0$          & $-2.0$          & $-7.0$          & $-7.0$          & dB    \\
      Beam radius on ground                         & $3.1$           & $3.1$           & $3.1$           & $3.1$           & km    \\
      Single-Beam throughput                        & $10.7$          & $10.7$          & $10.9$          & $10.9$          & Mbps  \\
      Number of active beams                        & $54$            & $72$            & $712$           & $1,785$         &       \\
      Aggregate throughput (active beams)           & $579$           & $772$           & $7,767$         & $19,473$        & Mbps  \\
      Coverage area                                 & $1,342$         & $1,789$         & $17,690$        & $44,351$        & km$^2$\\
      Area throughput $\rho$                        & $4.32$E$-1$     & $4.32$E$-1$     & $4.39$E$-1$     & $4.39$E$-1$     & Mbps/km$^2$\\ \hline
    \end{tabular}
    }
    \caption{Link budget and throughput estimation for the \ac{LEO} case.}
    \label{tab:LEO_System_parameters}
\end{table*}

Contrarily to the \ac{GEO} case, the size of the \ac{FoA} for a \ac{LEO} satellite is relatively small ($8.4\times8.4\,\textrm{m}^2$). In this case, a mechanically deployable large phased array with an aperture equivalent to that of our \ac{FoA} may represent an attractive alternative. On the other hand, envisaging a larger \ac{LEO} \ac{FoA} makes the size of each beam too small, creating exceedingly frequent satellite-induced handovers.  
\subsection{Performance results: \Acl{R-GEO} with winglets}\label{performance:winglets}
Finally, we investigate the potential advantage of introducing the winglets discussed in \sectionname~\ref{model:winglets}. To make a fair comparison, we studied three regional \acs{GEO} configurations with winglets of increasing size dubbed W1, W2, W3, retaining the same overall \ac{FoA} gain as in the case without winglets, named W0. The \acs{FoA} parameters are summarized in the first five lines of \tablename~\ref{tab:R-GEO_antenna_parameters_winglets}. Our main result is that we can keep the same \acs{FoA} directivity while  progressively reducing the number of satellites $S$ from $1089$ (reference case W0 without winglets) to $729$ for the case W1 (accommodating four winglets, each hosting one row of $7$ radiating elements), $529$ for the case W2 (accommodating four winglets, each hosting two rows of $7$ radiating elements each), and $441$ for the case W3 (accommodating four winglets, each hosting three rows of $7$ radiating elements each), respectively.

The corresponding system level performance results are summarized in the remaining lines of \tablename~\ref{tab:R-GEO_antenna_parameters_winglets}. We see that the introduction of the winglets allows us to increase the normalized system throughput while reducing the required number of satellites composing the \acs{FoA}. This is because the overall \acs{FoA}'s area  is increased by the presence of winglets despite the presence of ``holes''. The increased area also provides more overall DC/RF power, making the link budget improve.  In particular, configuration W2 provides a coverage similar to W0 with half the number of satellites,  a remarkable 58\% increase in total throughput and a 57\% improvement of normalized throughput per km$^2$. The configuration W3 allows to reduce by a factor $2.4$ the number of satellites at the expenses of a $32\%$ reduction in the coverage area but with a $11\%$ increase in the total throughput and a $62\%$ improvement of normalized throughput per km$^2$. Configuration W1 is the least attractive winglet configuration of the three, with severely reduced coverage area and throughput compared to W0.

These findings demonstrate that, despite the sparsity of the W2 \acs{FoA} geometry with associated increase in the antenna sidelobes level, the system performance is clearly much better than the configuration W0 without winglets. This comes at the expenses of the increased thickness of the single satellite arrays when folded for launch. 
\begin{table*}[h]
    \centering
    \small{
    \begin{tabular}{l c c c c c}\hline
      \emph{Parameter}                            & \emph{W0} & \emph{W1} & \emph{W2} & \emph{W3} & \emph{Unit}  \\\hline
      Number of array elements                      & $49$        & $49$        & $49$        & $49$        &       \\
      Number of radiating elements/winglet          & $0$         & $7$         & $14$        & $21$        &       \\
      Number of  satellites $S$                     & $1,089$     & $729$       & $529$       & $441$       &       \\
      Total \acs{FoA}     area                      & $14,762.25$ & $14,823.2$  & $14,342.0$  & $14,945.3$  & m$^2$ \\
      \acs{FoA} overall antenna directivity         & $70.33$     & $70.55$     & $70.50$     & $70.74$     & dBi   \\
      Single satellite total output \ac{RF} power   & $304.5$     & $424.3$     & $544.0$     & $663.8$     & W     \\
      \ac{FoA} overall output \ac{RF} power         & $54.3$      & $56.0$      & $57.0$      & $58.1$      & dBW   \\
      \ac{FoA} overall \ac{EIRP}/beam               & $85.0$      & $88.6$      & $85.9$      & $88.8$      & dBW   \\
      Line-of-sight downlink \ac{SNR}               & $13.6$      & $17.1$      & $14.5$      & $17.4$      & dB    \\
      Downlink \ac{SIR}                             & $12.9$      & $6.2$       & $8.4$       & $6.0$       & dB    \\
      Total \ac{SINR}  with shadowing               & $4.8$       & $4.0$       & $4.2$       & $4.0$       & dB    \\
      \acs{PHY} spectral efficiency                 & $0.82$      & $0.82$      & $0.82$      & $0.82$      & b/s/Hz\\
      Required \acs{PHY} \ac{SINR}                  & $0.0$       & $0.0$       & $0.0$       & $0.0$       & dB    \\
      Approximate beam radius on ground             & $14.2$      & $10.9$      & $11.3$      & $11.2$      & km    \\
      Single-beam throughput                        & $44.7$      & $44$        & $44.7$      & $44.7$      & Mbps  \\
      Number of active beams                        & $5,417$     & $3,684$     & $8,585$     & $5,946$     &       \\
      Aggregate throughput (active beams)           & $242.3$     & $163.2$     & $384.0$     & $265.9$     & Gbps  \\
      Coverage area                                 & $2,848,341$ & $1,122,709$ & $2,865,999$ & $1,928,531$ & km$^2$\\
      Area throughput $\rho$                        & $8.51$E$-2$ & $1.45$E$-1$ & $1.34$E$-1$ & $1.38$E$-1$ & Mbps/km$^2$\\ \hline
    \end{tabular}
    }
    \caption{Link budget and throughput estimation for the regional \ac{GEO} case T2-C1 with (W1, W2, W3) and without winglets (W0).}
    \label{tab:R-GEO_antenna_parameters_winglets}
\end{table*}
%

\section{System architecture and feeding/beamforming network}
\label{sec:implementation}
As already anticipated, the (good) results derived in the previous section can be actually experienced only if the \ac{FoA} is adequately fed by a wideband network/beamformer. The purpose of this section is to tackle this issue for the \ac{FoA} size of our study cases, and come up with possible solutions.

\subsection{System architecture}\label{feednet}

In line with the approach described in \cite{AST_patent2018}, we assume that the \ac{FoA}'s (small) array satellites are connected to a unique (larger) \ac{CS} (\figurename~\ref{fig:system_architecture}) performing the following tasks:
\begin{itemize}
\item {connect to the Earth \acp{GW} through several feeder links;} 
\item {perform feeder-link-to-beams routing, implementing  beamforming if this function is centralized;}
\item {connect the \ac{CS} to the \ac{FoA} satellites by means of \acp{ISL};}
\item {control the array satellites' relative positions;}
\item {compensate for carrier Doppler-shift and Doppler-rate affecting the center of the beams generated on-ground;}
\item {compensate for possible residual differential delay errors across the different arrays of the \ac{FoA}.}
\end{itemize} 
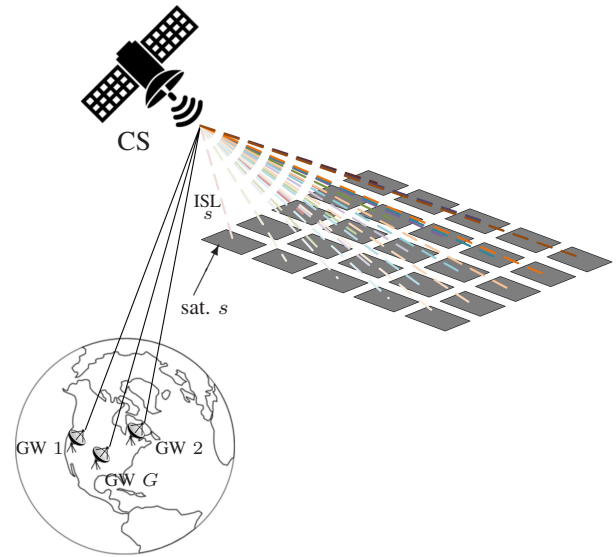
\begin{figure}[t]
  \begin{center}
    \begin{overpic}[width=\columnwidth]{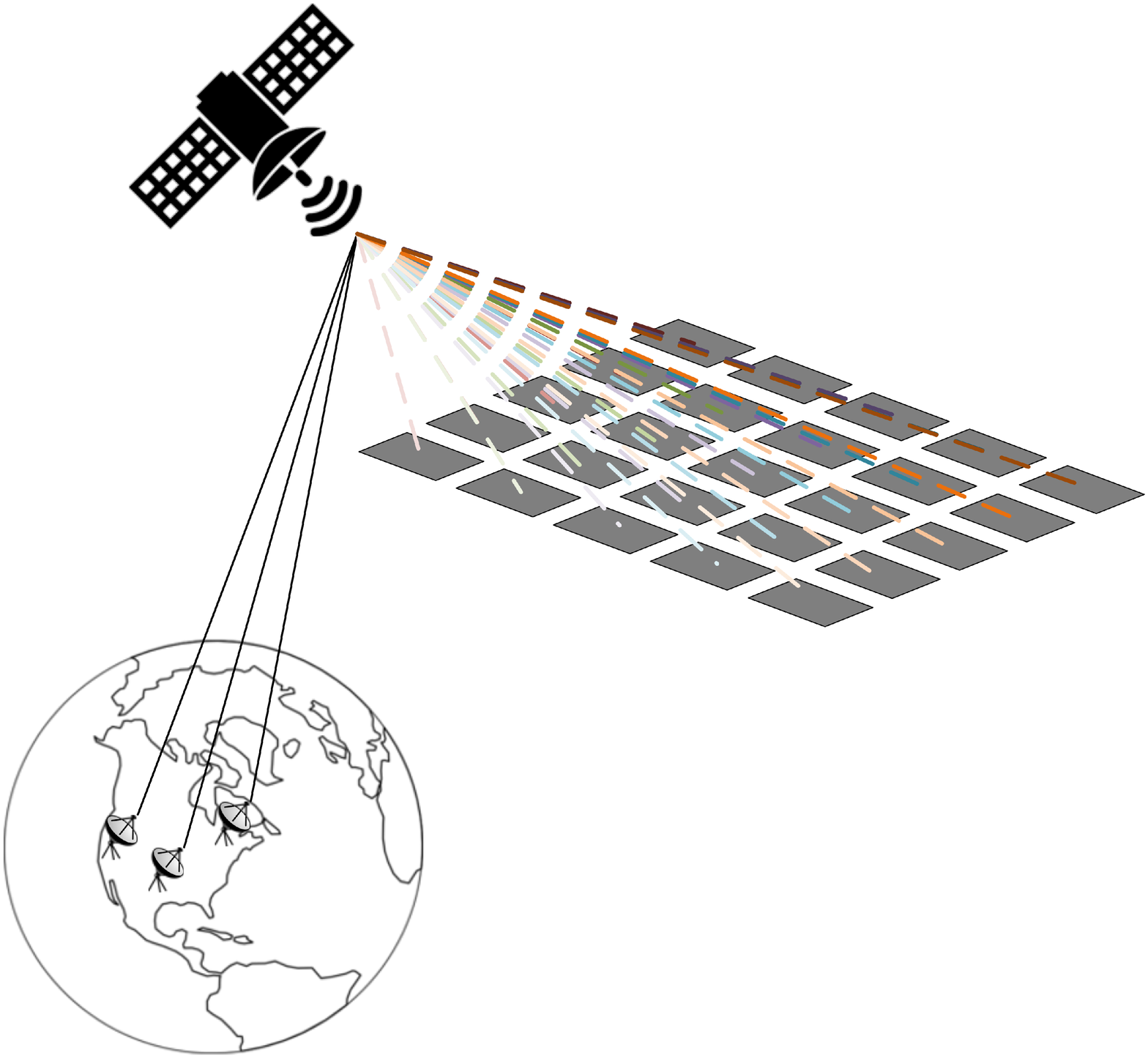}
      \put(06,22){\scriptsize{\acs{GW} $1$}}
      \put(28,22){\scriptsize{\acs{GW} $2$}}
      \put(20,17){\scriptsize{\acs{GW} $G$}}      
      \put(22,70){{\acs{CS}}}
      \put(34,47){\vector(1,2){4}}
      \put(32,44){\footnotesize{sat. $s$}}
      \put(34,61){\scriptsize{}{\acs{ISL}}}    
      \put(35.5,59){\scriptsize{}{$s$}}    
    \end{overpic} 
    \caption{{\acs{FoA} overall system architecture.}}
    \label{fig:system_architecture}
  \end{center}
\end{figure}

In our understanding, the \ac{CS} is a necessary entity for the \ac{FoA} as there is no possibility to directly feed the single satellites, because of the difficulty to address them individually from the Earth.

\begin{figure*}[t]
  \begin{center}
    \subfigure[\acs{CS} side.]{
    \begin{overpic}[height=0.24\textheight]{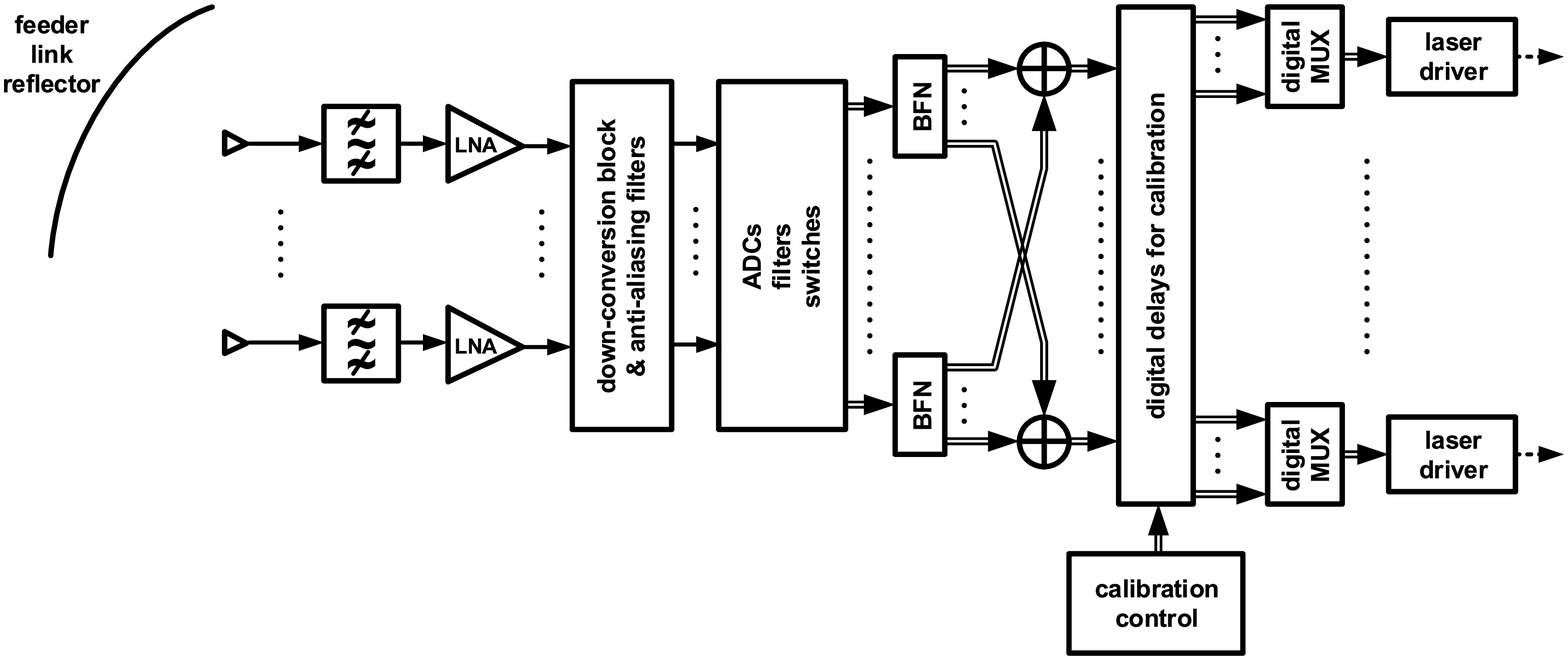}
      \put(17,34){\scriptsize{$1$}}
      \put(16,17.5){\scriptsize{$G$}}
      \put(43.5,34){\scriptsize{$1$}}
      \put(43.5,17){\scriptsize{$G$}}
      \put(54.5,36.5){\scriptsize{$1$}}
      \put(54.5,13.5){\scriptsize{$K$}}
      \put(60.5,39){\scriptsize{$1$}}
      \put(60.5,30.5){\scriptsize{$S\!N$}}
      \put(60.5,19.5){\scriptsize{$1$}}
      \put(60.5,11.5){\scriptsize{$S\!N$}}
      \put(69,39){\scriptsize{$1$}}
      \put(68.5,11.5){\scriptsize{$S\!N$}}
      \put(77.2,42){\scriptsize{$1$}}
      \put(77,33.5){\scriptsize{$N$}}
      \put(77.2,16){\scriptsize{$1$}}
      \put(77,08){\scriptsize{$N$}}
      \put(86.7,39.5){\scriptsize{$1$}}
      \put(86.5,10.5){\scriptsize{$S$}}
      \put(98,40){\scriptsize{\acs{ISL} $1$}}
      \put(98,14.5){\scriptsize{\acs{ISL} $S$}}
    \end{overpic}
    \label{fig:centralizedBFN_tx}}
    \\
    \subfigure[array satellite side ($s$th satellite).]{
    \begin{overpic}[height=0.08\textheight]{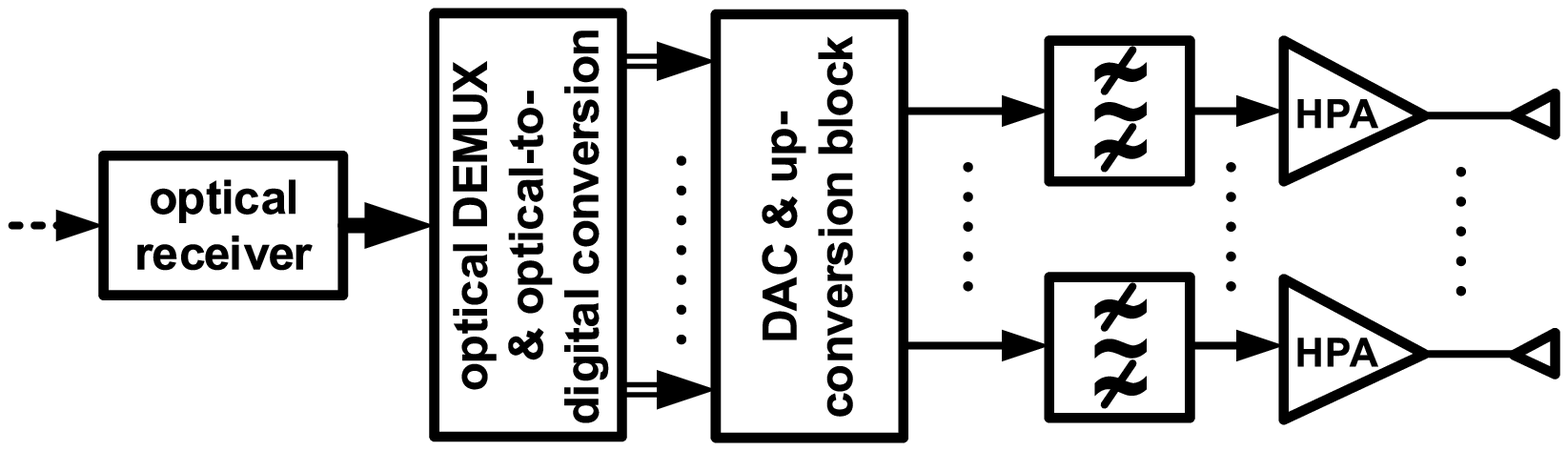}
      \put(-4,18){\scriptsize{\acs{ISL} $s$}}
      \put(40.5,27){\scriptsize{$1$}}
      \put(40,-1){\scriptsize{$N$}}
      \put(59.5,24){\scriptsize{$1$}}
      \put(59,1.3){\scriptsize{$N$}}
      \put(92.3,24){\scriptsize{$1$}}
      \put(92,1.3){\scriptsize{$N$}}
    \end{overpic}    
    \label{fig:centralizedBFN_rx}}
    \caption{{Centralized \acs{BFN} payload architecture.}}
    \label{fig:centralizedBFN}
  \end{center}
\end{figure*}

\subsection{Beamforming/feeding network}
{The sparsity of our \ac{FoA} antenna architecture allows for two distinct approaches in the \ac{BFN} implementation, namely, distributed and centralized.}

{At a first glance, the \ac{BFN} can be more easily implemented by each single satellite in a distributed fashion. In this option, the \ac{CS}  splits the feeder-link beam signals coming from the \acp{GW}, and routes them to the appropriate individual satellite(s) by means of a dedicated, local \ac{ISL}. Each array satellite will individually take care of the \ac{BFN} implementation. The other option is to have all \ac{BFN} functions centrally implemented by the \ac{CS}, and the relevant pre-processed outputs sent to the individual satellites via the same \acp{ISL}.}

{It is easy to see that the distributed \ac{BFN} architecture has a major drawback for the most demanding \ac{R-GEO} case. This is because the number of active beams $K$ is much larger than the number of the satellites' array elements $N$. As a consequence, the throughput of each \ac{ISL} (proportional to $K$) is much higher than in the case of the centralized \ac{BFN} architecture (throughput proportional to $N$). For this reason,  we will focus in the following on centralized beamforming only.} It should be remarked that a true delay \ac{BFN} should be adopted for the \ac{CS} when the narrowband phased array condition \cite{Mailloux_book} is not verified. This corresponds to the case when the signal bandwidth to carrier frequency ratio $B/f_0$ does not satisfy the condition
\begin{align}
\frac{B}{f_0} \ll \frac{\lambda}{D\sin\left(\theta_{\rm max}\right)},
\end{align}
where $D/\lambda$ is the normalized \ac{FoA} dimension, and $\theta_{\rm max}$ is the maximum array scan angle. {\figurename~\ref{fig:centralizedBFN} depicts a sample architecture of a centralized \ac{BFN} for our \ac{FoA}. The  $G$ incoming feeder links to the CS are  filtered, amplified and down-converted to an adequate intermediate frequency that is compatible with the digital processor  front-end. The feeder links operating at Ka-band are all reusing the same frequency. This is made possible by exploiting  spatial  separation among \acp{GW} and sufficiently directive \ac{CS} receive antennas. Assuming standard feeder-link \ac{RF} bandwidth, the overall feeder throughput may require a large number of Earth \acp{GW}.}

{Considering the total (large) number of \acp{FoA} elements and the (large) number of beams  to be generated, the best solution to implement the \ac{BFN} is a so-called \textit{transparent} digital processor. The transparent processor just digitizes all of the input analog signals with adequate conversion frequency, and then  takes care of routing the (digital) feeder link signals to the desired user link beam, after filtering, switching and beamforming. The output of the $S\times N$ true delay \acp{BFN} are multiplexed $N$ at a time to generate the $S$ individual digital streams to be sent through the respective \acp{ISL} to the  different array satellites. In the schematic of \figurename~\ref{fig:centralizedBFN} we just refer for simplicity to a single-color configuration as in the T2 case.}

We drafted the specifications of the feeder link, the \ac{ISL}, and the digital processor for the most demanding \ac{R-GEO} and \ac{LEO} cases discussed in \sectionname~\ref{sec:results} -- the relevant results  are shown in \tablename~\ref{tab:ISL+FL+Processor_sizing}. As can be noticed, the total \ac{ISL} bandwidth required to connected the CS to each satellite  is not compatible with an \ac{RF}-based solution, and this is why we envisaged an  optical \ac{ISL} from the very design onset. In particular, optical digital modulation is certainly preferable to simplify the digital-to-optical \ac{ISL} section interface and to prevent possible calibration issues.

{For the \ac{LEO} case, the \ac{ISL} bit rate is higher than for the \ac{R-GEO} since the number of feeds per satellite is $45$ times higher. In this case, \ac{WDM} on each single \ac{ISL} will be required. The \ac{ISL} digital optical message is received on each satellite and then demultiplexed into $N$ digital streams corresponding to the different $N$ array-element signals. After \ac{DAC} and up-conversion to the downlink \ac{RF} frequency, the $N$ RF chains are filtered, amplified and sent to the $N$ array elements.}

{As indicated in \figurename~\ref{fig:centralizedBFN}, the final \ac{DAC}-upconversion-array-element feeds need to be calibrated by digitally pre-correcting the array differential delay errors.} Fortunately, the \ac{CS} tight control of the \ac{FoA}'s geometry allows to limit the differential path delays, reducing the residual calibration to a (centralized) fine delay control. Possible solutions to derive the calibration data about the array \ac{RF} chains are  detailed in \sectionname~\ref{subsec:calibration}.

\begin{table*}[h]
{
    \centering
    \small{
    \begin{tabular}{l c c c}\hline
      \emph{Parameter}                        & \emph{(\acs{R-GEO}) T2, C2} & \emph{(\acs{LEO}) T2, C2} & \emph{Unit}   \\ \hline
      Number of beams   $K$                     & $5,417$                       & $1,785$                     &        \\
      Number of colors  $C$                     & $1$                           & $1$                         &        \\
      Number of array elements  $N$             & $49$                          & $2,209$                     &        \\
      Number of satellites $N$                  & $1,089$                       & $5$                         &        \\
      Bandwidth/beam                            & $60$                          & $60$                        & MHz    \\
      Number of bits \acs{ADC}/\acs{DAC}        & $8$                           & $8$                         &        \\
      Number of samples/bandwidth               & $2.5$                         & $2.5$                       &        \\
      Optical modulation spectral efficiency    & $2$                           & $2$                         & bits/symb\\
      Single \acs{ISL} throughput \acs{BFN}     & $59$                          & $2,651$                     & Gbps    \\
      Occupied ISL bandwidth                    & $59$                          & $2,651$                     & GHz     \\
      Number of \acs{WDM} required per \acs{ISL}& $1$                           & $27$                        &        \\
      Aggregated \acs{ISL} bandwidth            & $64,251$                      & $13,255$                    & GHz    \\
      Feeder link total bandwidth               & $325.02$                      & $107.1$                     & GHz    \\
      Available feeder link \ac{RF} bandwidth   & $3$                           & $3$                         & GHz    \\
      Number of feeder link polarizations       & $2$                           & $2$                         &        \\
      Number of gateways required for
      each \ac{FoA} $G$                         & $55$                          & $18$                        &        \\\hline
    \end{tabular}
    }
    \caption{{\acs{FoA} Feeder link, \acs{ISL} and processor sizing for the most demanding \ac{R-GEO} and \ac{LEO} cases.}}
    \label{tab:ISL+FL+Processor_sizing}
    }
\end{table*}
\subsection{{Calibration issues}}
\label{subsec:calibration}
{We previously analyzed the impact of the array feeding chains instability upon the \ac{FoA} resulting beam pattern. The possible differential phase errors are caused by \ac{RF} chain errors or uncompensated \ac{FoA} geometrical instabilities.}

Considering stability of the formation, a very good overview of satellite \ac{FF} technology and missions is provided in \cite{FF_control_survey}, wherein modern control techniques are in particular discussed. A summary of the performance of NASA's \ac{FF} missions for remote sensing, associated technologies and \ac{GNC} is also reported in \cite{NASA_FF}.  A more recent detailed overview of the current and future \ac{GNC} and \ac{FF} solutions is contained in \cite{FF_GNC_Di_Mauro}. The most typical \ac{GNC} techniques are based on \ac{GNSS} accurate positioning complemented by radio frequency or optical satellite-to-satellite metrology. Our \ac{FF} relative satellite position accuracy requirement is in the order of $1$ cm (i.e., $1/10$ of the wavelength, see \cite{Bekey2006} and \figurename~\ref{fig:2d_spacing}) at a distance of a few meters, closely matching the performance required by scientific missions \cite{NASA_FF,FF_GNC_Di_Mauro}. In addition, this stringent \ac{FF} relative position accuracy has to be implemented in relatively small and low-cost satellites.  The control accuracy challenge is even more striking for \ac{LEO}  than for \ac{GEO} orbits considering the larger differential  acceleration affecting the spacecrafts. Fortunately, the \ac{GNC} challenge can be mitigated in our case considering that the (relative) position error affecting each satellite can be compensated, as far as beamforming is concerned, by \emph{real-time calibration} of the \acp{FoA}: the carrier phase of each (inter-satellite) link from the master satellite feeding the \acp{FoA} to the other \acp{FoA} satellites can be adjusted so as to re-phase the different elements. This means that the \emph{cm}-level accuracy requirement applies to the accuracy of position \textit{determination} rather than to position \textit{keeping}. The patent \cite{AST_patent2018} discusses an approach to control the relative position of the array satellites using the force generated by electromagnetic coils on top of gravitational ones. The array satellite instantaneous position is based on a \ac{GNSS} receiver installed on-board that provides measurements to control the electromagnetic actuators. The conclusion from {the review of} the references above is that an \ac{FoA} is challenging but realizable with the best of current technology, but that some degree of instability has anyway to be accounted for.

A good overview of the calibration aspects for active antennas is contained in \cite{Angeletti_calibration}. The selection of the most appropriate calibration solution for the proposed \acp{FoA} is beyond the scope of the current paper.

\section{Conclusions} 
\label{sec:ConcSect}
In this paper, we have presented a theoretical analysis of an \ac{FoA} antenna, its preliminary design optimization, and we have shown how to take advantage of this new technology to improve network throughput in a multi-beam S-band mobile communication system  with low-earth or geostationary orbiting satellites. The goal was on one hand to prove that 4G-like communication services to hand-held user terminals are indeed feasible, but on the other that the adoption of new technologies are necessary.

In particular, we have shown that the \ac{FoA} technique, albeit challenging, appears feasible with state-of-the-art satellite and antenna technologies. Whether it is demonstrated that calibration errors among the sub-arrays do not appear to have a major performance impact, the formation flying requirements are very demanding. Through an appropriate system analysis methodology that was specifically developed, we have shown that a very large \ac{FoA} ($15,000$ m$^2$) in a geostationary orbit can provide almost continental coverage with a network throughput $\rho$ of $0.02$ to $0.1$ Mbps/km$^2$, comparable to that of a rural terrestrial cell.  For the \ac{LEO} scenario, the \ac{FoA} solution appears less attractive than for \ac{GEO} systems because the necessary antenna aperture to be synthesized is way smaller than that necessary for the \ac{GEO} orbit. Formation-flying-based antennas for \acp{LEO} are therefore an interesting option but not strictly necessary, as mechanically deployable phased arrays may represent an attractive alternative. 

Further work is needed to optimize the radiating pattern of the \ac{FoA} through appropriate optimization techniques, in order to decrease inter-beam interference and increase network throughput. The way to achieve the required formation flying accuracy, as well the payload technological assessment and the satellite power sub-system accurate modelling, are other aspects requiring further investigation.

\section*{Acknowledgment} 
\addcontentsline{toc}{section}{Acknowledgment}
The authors would like to thank Dr. Piero Angeletti, from the European Space Agency (ESA), for his contributions and insightful discussions on aspects related to the active antenna and payload design aspects. A special thank goes to Dr. Miguel {\'A}. Vazquez from the Centre Tecnol{\`o}gic de Telecomunicacions de Catalunya (CTTC) for his remark on the applicability of the narrowband phased-array model.

\bibliographystyle{IEEEtran}
\bibliography{IEEEabrv,refs,FoA_v3}

\begin{IEEEbiography}
[{\includegraphics[width=1.0in,height=1.25in,clip,keepaspectratio]{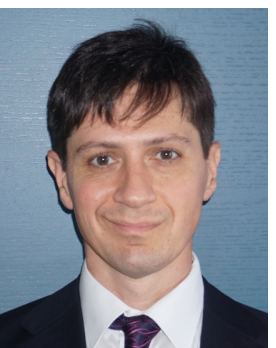}}]
{Giacomo Bacci} (M'09) received the B.E. and M.E. degrees in telecommunications engineering and the Ph.D. degree in information engineering from the University of Pisa, Pisa, Italy, in 2002, 2004, and 2008, respectively. From 2006 to 2007, he was a visiting student research collaborator with Princeton University, Princeton, NJ, USA. From 2008 to 2014, he was a post-doctoral research fellow with the University of Pisa. From 2008 to 2012, he was also a software engineer with Wiser Srl, Livorno, Italy, and from 2012 to 2014, he was also enrolled as a visiting post-doctoral research associate with Princeton University. From 2015 to 2021, he was a product manager for interactive satellite broadband communications at MBI Srl, Pisa, Italy. Since 2022, he joined the University of Pisa as a tenure-track assistant professor.

Dr. Bacci is the recipient of the FP7 Marie Curie International Outgoing Fellowships for career development (IOF) 2011 GRAND-CRU, the Best Paper Award from the IEEE Wireless Communications and Networking Conference (WCNC) in 2013, the Best Student Paper Award from the International Waveform Diversity and Design Conference (WDD) in 2007, the Best Session Paper at the ESA Workshop on EGNOS Performance and Applications in 2005, and the 2014 URSI Young Scientist Award. He was named as an Exemplary Reviewer 2012 for IEEE Wireless Communications Letters.
\end{IEEEbiography}

\begin{IEEEbiography}
[{\includegraphics[width=1.0in,height=1.25in,clip,keepaspectratio]{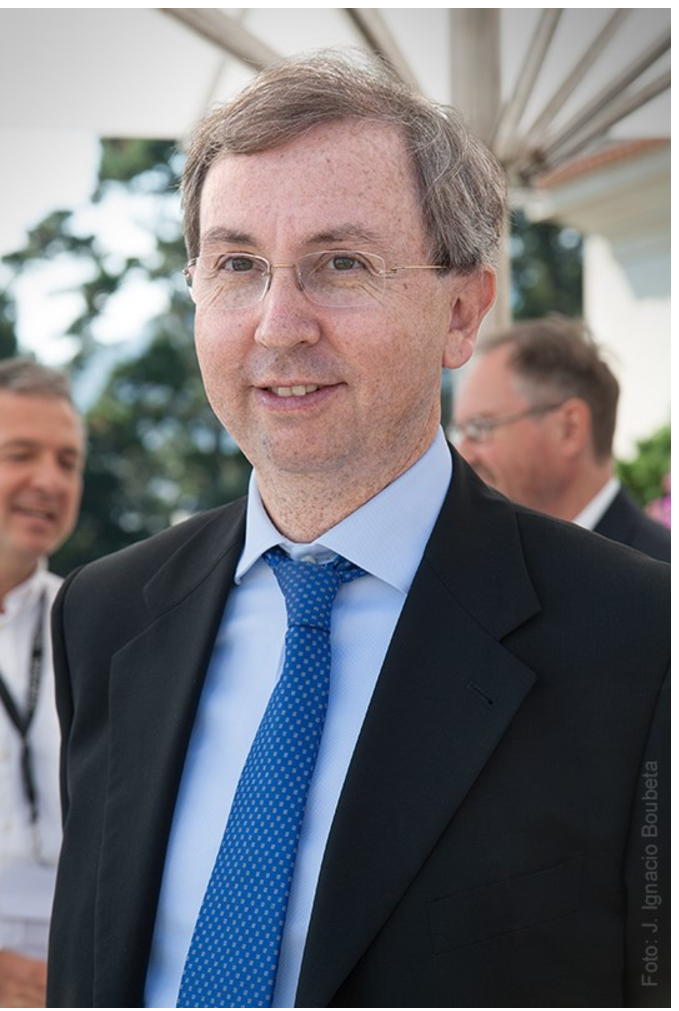}}]
{Riccardo De Gaudenzi} (M'88 - SM'97 - F'22) received his Doctor Engineer degree (cum Laude) in electronic engineering from the University of Pisa, Italy in 1985, a PhD in Code Division Multiple Access for Personal Communication Systems from the Technical University of Delft, The Netherlands in 1999, and a Communication Engineering Master Degree ad honorem from University of Parma in 2021. From 1986 to 1988 he was with the European Space Agency (ESA), Stations and Communications Engineering Department, Darmstadt (Germany) where he was involved in satellite Telemetry, Tracking and Control (TT\&C) ground systems design and testing. In 1988, he joined ESA's Research and Technology Centre (ESTEC), Noordwijk, The Netherlands, where he has been covering several technical and managerial positions inside the Directorate of Technology, Engineering and Quality. He is currently the Head of the ESA's Electrical Engineering Department. He has led a large number of R\&D activities for TT\&C, Telecom and Navigation applications. In 1996 he spent one year with Qualcomm Inc., San Diego USA, in the Globalstar LEO project system group under an ESA fellowship.

His current interest is mainly related with efficient digital modulation and multiple access techniques for fixed and mobile satellite services, synchronization topics, adaptive interference mitigation techniques and communication systems simulation techniques. He actively contributed to the development and the demonstration of the ETSI S-UMTS Family A, S-MIM, DVB-S2, DVB-S2X, DVB-RCS2 and DVB-SH standards.  He has published more than 140 scientific papers and owns more than 30 patents. From 2001 to 2005 he has been serving as Associate Editor for CDMA and Synchronization for IEEE Transactions on Communications and Associate Editor for Journal of Communications and Networks. He is co-recipient of the 2003 and 2008 Jack Neubauer Memorial Award Best Paper from the IEEE Vehicular Technology Society. He was awarded the AIAA 2022 Aerospace Communications Award.
\end{IEEEbiography}

\begin{IEEEbiography}
[{\includegraphics[width=1.0in,height=1.25in,clip,keepaspectratio]{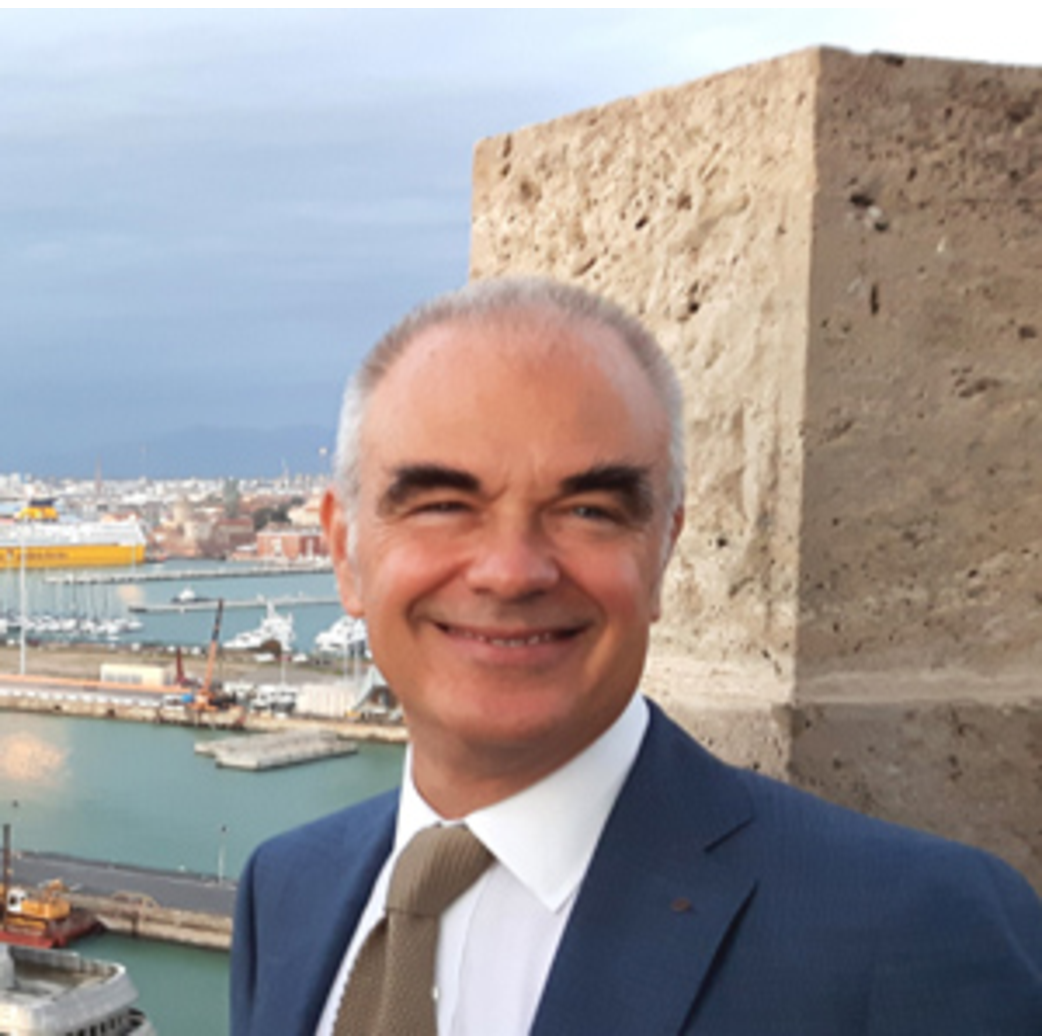}}]
{Marco Luise}(M' - SM'- F') is a Professor of Telecommunications at the University of Pisa and at the University of Florence, Italy, and was in the past a Research Fellow of the European Space Agency. He has chaired a number of scientific conferences, including IEEE ICASSP 2014 in Florence, Italy and IEEE ICC 2023 in Rome, Italy. A former Editor of the IEEE Trans. Communications, he is a Division Editor of the J. Commun. and Networks, and was the coordinator of the European Network of Excellence in Wireless Communications NEWCOM\#. An IEEE Fellow, he has authored more than 300 publications, and his main research interests lie in the broad area of wireless/satellite communications and positioning.
\end{IEEEbiography}

\begin{IEEEbiography}
[{\includegraphics[width=1.0in,height=1.25in,clip,keepaspectratio]{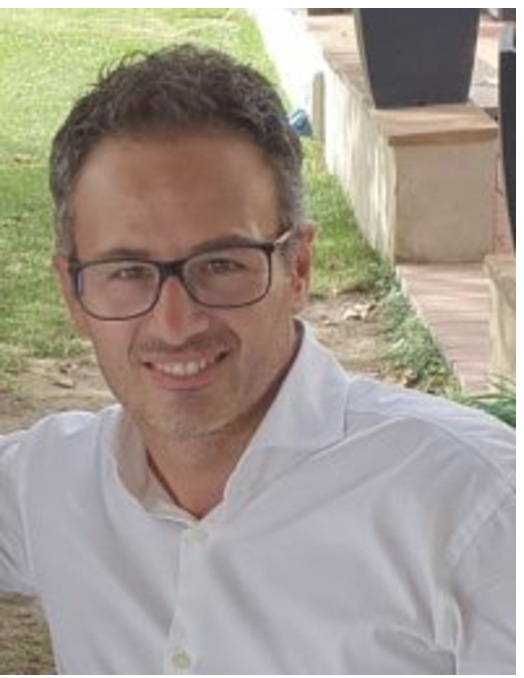}}]
{Luca Sanguinetti}(M'06 - SM'15) received the Laurea Telecommunications Engineer degree (cum laude) and the Ph.D. degree in information engineering from the University of Pisa, Italy, in 2002 and 2005, respectively. In 2004, he was a visiting Ph.D. student at the German Aerospace Center (DLR), Oberpfaffenhofen, Germany. During the period June 2007 - June 2008, he was a postdoctoral associate in the Dept. Electrical Engineering at Princeton. From July 2013 to October 2017 he was with Large Systems and Networks Group (LANEAS), CentraleSup\'elec, France. He is currently an associate professor in the ``Dipartimento di Ingegneria dell'Informazione'' of the University of Pisa, Italy.

He served as an associate editor for IEEE Trans. Wireless Communications and IEEE Signal Processing Letters, and as lead guest editor of IEEE J. Selected Areas of Communications, Special Issue on ``Game Theory for Networks'', and as an associate editor for IEEE J. Selected Areas of Communications, series on Green Communications and Networking. Dr. Sanguinetti is currently serving as an associate editor for the IEEE Trans. Communications and is a member of the Executive Editorial Committee of IEEE Trans. Wireless Communications.

His expertise and general interests span the areas of communications and signal processing. Dr. Sanguinetti co-authored the textbooks \emph{Massive MIMO Networks: Spectral, Energy, and Hardware Efficiency} (2017) and \emph{Foundations on User-centric Cell-free Massive MIMO} (2020). He received the \emph{2018} and \emph{2022 Marconi Prize Paper Award in Wireless Communications} and co-authored a paper that received the young best paper award from the ComSoc/VTS Italy Section. He was the co-recipient of two best conference paper awards: \emph{IEEE WCNC 2013} and \emph{IEEE WCNC 2014}. He was the recipient of the FP7 Marie Curie IEF 2013 ``Dense deployments for green cellular networks''. 
\end{IEEEbiography}

\begin{IEEEbiography}
[{\includegraphics[width=1.0in,height=1.25in,clip,keepaspectratio]{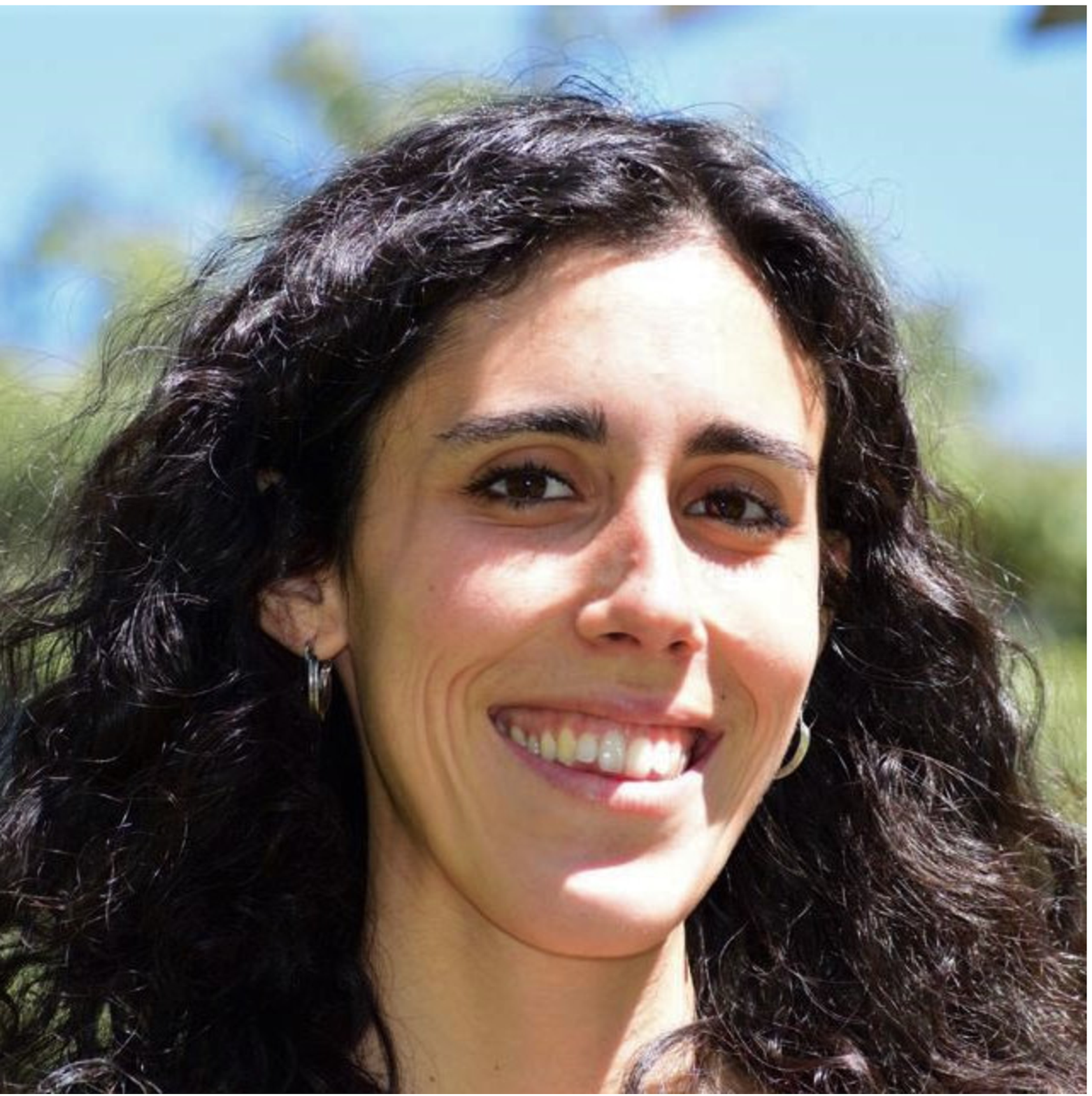}}]
  {Elena Sebastiani} received the B.E. and the M.E. degrees in aerospace engineering in 2017 and 2020, respectively, from the University of Pisa, Pisa, Italy, where she is currently working toward the Ph.D. degree in information engineering. Since 2021, she has been with Wiser Srl, Livorno, Italy.
  
Her research interests are in the areas of satellite payload, antenna design for space applications, and satellite constellations.
\end{IEEEbiography}%

\end{document}